\documentclass{article}

\usepackage{arxiv}

\usepackage[utf8]{inputenc} % allow utf-8 input
\usepackage[T1]{fontenc}    % use 8-bit T1 fonts
\usepackage{hyperref}       % hyperlinks
\usepackage{url}            % simple URL typesetting
\usepackage{booktabs}       % professional-quality tables
\usepackage{amsfonts}       % blackboard math symbols
\usepackage{nicefrac}       % compact symbols for 1/2, etc.
\usepackage{microtype}      % microtypography
\usepackage{lipsum}		% Can be removed after putting your text content
\usepackage{graphicx}
\usepackage{natbib}
\usepackage{doi}
\usepackage{bbm}
\usepackage[table]{xcolor}
\usepackage{multirow}
\usepackage[normalem]{ulem}

\usepackage[american]{babel}
\usepackage{latexsym,bm}
\usepackage{amsmath,amssymb}
\usepackage{amsthm}
\usepackage{csquotes}

\title{Stereotype graph: A mathematical framework of category stereotypes via graph theory}

\author{ \href{https://orcid.org/0000-0003-2683-6095}{\includegraphics[scale=0.06]{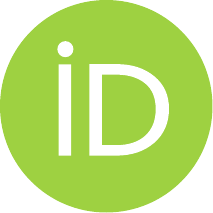}\hspace{1mm}Yijia Yan}\thanks{This paper was finished and presented in 2016 for the 8th Science Innovation of the School of Mathematical Sciences at Fudan University, Shanghai, China.} \\
	Nuffield Department of Clinical Neurosciences\\
    Wellcome Centre for Integrative Neuroimaging\\
	University of Oxford\\
	Oxford, United Kingdom \\
	\texttt{yijia.yan@ndcn.ox.ac.uk} \\
}

\date{} 

\renewcommand{\shorttitle}{Category stereotype graph}

\hypersetup{
pdftitle={Stereotype graph: A mathematical framework of category stereotypes via graph theory},
pdfauthor={Yijia Yan},
pdfkeywords={Graph theory, Category stereotype, Vertex coloring, Stereotype graph, Chromatic stability index (CSI)},
}

\begin{document}
\maketitle

\begin{abstract}
In social psychology and cognitive science, there has been much interest in studying category stereotypes. However, we still lack a consensual mathematical definition or framework, which is necessary for us to hold a deeper understanding of stereotypes in human cognition. In this paper, we use graph theory to portray category stereotypes in human cognition, based on pairs of labels having special relations. By using methods and conclusions in graph theory (including algebraic graph theory and vertex coloring) as well as strict ratiocination, we give criteria for judging the stability of a given stereotype, some of which are computationally practicable. We also define the chromatic stability index (CSI) to measure the stability of a stereotype in human cognition, as well as to provide its precise range. From the perspective of stereotype graphs and CSI, we may explain why stereotypes can easily stay in human cognition.
\end{abstract}


\section{Introduction}
      \emph{Stereotype}, a terminology in social psychology and cognition, is any adopted belief or thought about specific types of individuals or behaviors, as well as representing them as a whole \citep{McGarty2002}. Stereotype in this article refers to \emph{category stereotype}, which is a \emph{personal} cognitive belief about judging people as category members rather than individuals \citep{Perry2000a}. Most researchers use the word \emph{stereotype} directly to refer to category stereotypes. We use both of terminologies interchangeably in this article.

      It is commonly believed that \citet{Lippmann1922} was one of the first study of stereotypes in the modern psychological sense. Early cognitive approaches to stereotypes may trace to decades of years ago \citep{Lippmann1922,Allport1954}. Although stereotypes has been studied for a long time, there are very few models of stereotypes and stereotyping. In contrast, the stereotype content model \citep{Fiske2002} provides a simple framework for stereotype content. In recent years, there is a growing body of researches about stereotypes in psychology and cognitive science \citep{Song2016,Hinzman2017,Lindqvist2017}.

      However, we still lack a consensual mathematical framework of category stereotypes. Such a framework would enable us to understand stereotypes better in human's cognition, and analyze stereotypes computationally. Moreover, such a theoretical framework would provide a brand-new perspective and thought of studying stereotypes and even in the field of social psychology, as well as hold the potential for being applied to other studies. In this article, we attempt to build a mathematical framework via graph theory and vertex coloring.

      The category is called \emph{label}, or \emph{characteristic}. We use both of these terminologies interchangeably, although we mainly use \emph{label} in this article. Label includes not only traits (e.g. extrovert, femininity), but also physical characteristics (e.g. male, big feet), social status (e.g. wealthy person, statesman), etc.
      
      Besides, relations (or linkage) among labels also decide how we categorize others with stereotypes. Three kinds of relations are listed below, which are the core of our mathematical framework in this article.

      \begin{description}
	    \item[Claim]
        Generally speaking, we consider a pair of labels presented by Label 1 and Label 2 in one's cognition, respectively.

        If no one can be categorized by Label 1 or Label 2, then Label 1 and Label 2 are \emph{contradictory} and they are a pair of \emph{contradictory labels}.

        If everyone can be categorized to either Label 1 or Label 2, then Label 1 and Label 2 are \emph{complete} and they are a pair of \emph{complete labels}.

        A pair of labels which is both contradictory and complete is called a pair of \emph{complementary labels}, and they are \emph{complementary}.
      \end{description}

      For example, Anne may regard anyone as either a beneficent person or a severe person. Here \emph{Beneficence} and \emph{Severity} can be regarded as two labels in Anne's cognition. In other words, Anne thinks that no one would be both beneficent and severe, because they're contradictory. Then we can say, in Anne's cognition, ``Beneficence'' and ``Severity'' is a pair of \textit{contradictory labels}. ``Beneficence'' and ``Severity'' are \textit{contradictory}.

      Another example is that Anne thinks anyone would be either beneficent or severe. Then we say that in Anne's cognition, ``Beneficence'' and ``Severity'' is a pair of \textit{complete labels}. ``Beneficence'' and ``Severity'' are \textit{complete}.

      No matter what labels occur in Anne's cognition, it can be viewed as a stereotype. To simplify the description, we call such labels \emph{stereotype-related labels}. They include contradictory labels, complete labels, and complementary labels.

      Note that the stereotype is actually \emph{personal}, which means different people may hold different stereotypes. This idea is based on implicit personality theory \citep{Bruner1954}. It's also worth mentioning that the stereotype doesn't always have to be activated. That is to say, the stereotype doesn't always take part in a cognitive process. Even activated, it may not be used as the basis of impression formation or judgments \citep{Marilynn1996}.

      Illustrating the relationship among multiple stereotype-related labels becomes quite challenging with a large number of labels. We use graph theory, especially vertex coloring, to construct a mathematical framework as an approach.

    \subsection{Graph Theory and Stereotypes Graphs}
      As a fundamental tool in mathematics and computer science, graph theory has been applied to psychology for decades \citep{Riley1969}. Recently, it has been applied to various fields, including neuroscience \citep{Xu2014,Brier2014,Vecchio2016,Vecchio2017}, social psychology \citep{Grandjean2016}, cognitive science \citep{Grady2015,Kellermann2015}, etc.

      In this paper, we use graph theory to describe various stereotypes based on stereotype-related labels. We regard labels as vertices, while relations among labels as the adjacency of vertices (i.e. edges). Every such \textit{finite simple graph} (we call it \textit{graph} briefly afterwards) we achieve is called a \textit{stereotype graph}, which is the base of our mathematical framework. We assume that readers are already familiar with basic concepts of graph theory.

      How does a stereotype graph relate to cognitive thinking with stereotypes? We make a quick review on three important components of a stereotype \citep{Perry2000a}:
      \begin{enumerate}
        \item A group of people are identified by a specific characteristic.
        \item We then attribute a set of additional characteristics to the group as a whole.
        \item Finally, on identifying a person as having the identifying meaningful characteristic, we then attribute the stereotypical characteristic to them.
      \end{enumerate}

      The first component is used to build specific labels as \emph{vertices}. The second component refers to building relations among labels as \emph{edges}. The third component is a \emph{stereotyping process}, through which the stereotype is activated to judge or categorize people. Such a stereotyping process is used to build a quantified index for measuring how stable can a stereotype be. In this way, the stereotype graph represents related cognitive thinking which may be considered as a stereotype. Readers may see more related expositions in the corresponding section.

      After we introduce several basic thoughts, conclusions, and terminologies, we give the formal definition of stereotype graphs with several clear and general statements about the correspondence between the stereotype graph and the category stereotype.

      In addition, we assume that for every label, there always exist someone who can be categorized into it. This ensures labels not be meaningless in the stereotype graph.

      The meaning of a stereotype graph can change completely when we consider different kinds of relations among labels. We first consider complementary labels, then contradictory labels. Other types of relations may be similar.

    \subsection{Stability of Stereotype Graphs}
      The stereotype is not likely to change when facing counter-stereotypical information, and the change appears to depend very much on the circumstances \citep{Perry2000b}. The stereotype may remain stable with the help of some strategies, including self-confirming, stereotype preservation biases and subtyping \citep{Perry2000b}. We mention it again from the perspective of stereotype graphs in the discussion section.

      Nonetheless, the stereotype may still change when someone perceives enough individual information of a certain person \citep{Marilynn1996}. For instance, one can detect paradoxes unavoidably from the stereotype when perceiving enough detailed information. We raise two examples of such paradoxes, named \emph{logical errors} and \emph{inevitable errors} respectively. We assume that the stereotype has to be changed under these circumstances. It is important to determine whether a stereotype is subject to change or not. This crucial property of stereotype graphs is called \emph{stability}. Readers may get a better sense of it with the perspective of stereotype graphs.

      We first discuss stereotype graphs with complementary labels and their stability. With the help of merge operations, we say the graph is \textit{bipartitely stable}  when it has no logical errors. After explicit discussion, we can show whether the graph is bipartitely stable through several mathematical criteria summarized in a specific section. Many of these criteria are easy to use computationally, which makes it more convenient to judge the stability of a stereotype graph. We also give related proofs of each criterion.

      We also discuss stereotype graphs with contradictory labels, which contain those with complementary labels. Since the labels no longer become complementary, we then discover the stability by considering another inevitable error based on the stereotyping process and stereotype preservation biases. The stereotype preservation biases refers to the tendency that we prefer confirming information to disconfirming information when we test our stereotypical beliefs \citep{Johnson1996}.

      In terms of vertex coloring, we then define chromatic stability index (or CSI briefly) as the chromatic number of the stereotype graph to quantify how stable a stereotype graph can be. After formal proofs, we know the exact range of the CSI is every integer in $[2,n]$ where $n\ge2$ is the number of pairs of labels. Moreover, bipartite graph $K_{n,n}$ and complete ladder graph $KL_{n}$ are the unique graph of the stereotype graph whose CSI is 2 and $n$ respectively.\footnote{Explanations of corresponding notations and proofs are given when we later discuss stereotype graphs based on contradictory labels}

      Finally, we discuss several further topics on stereotype graphs. We can extend the stereotype graph to more usual ones by deleting some edges (i.e. abandoning a part of the stereotype), while it won't decrease the stability. Since it seems ``stable'', it is, to some extent, a manifestation of the stereotype effect and the resistance of stereotypes to change.

      Without describing especially, we borrow some standard notations and terminologies from the book ``Modern Graph Theory'' \citep{Bollobas1998} and the book ``Algebraic Graph Theory'' \citep{Biggs1974}. Of course, we describe some important notations the first time they occur in this article.

  \section{Stereotype Graphs based on Complementary Labels}
      We give a more complicated example of stereotype, depending on the relation between two pairs of complementary labels. Suppose that one of the pair of labels is ``Beneficence'' and ``Severity'', as Anne would say:`` If someone isn't beneficent, then he must be severe, and vice versa.'' The other pair that Anne considers is ``Femininity'' and ``Masculinity'', as Anne would say:`` If someone isn't feminine, then he must be masculine, and vice versa.''
      \begin{figure}
      	\centering
      	\includegraphics[height=4cm]{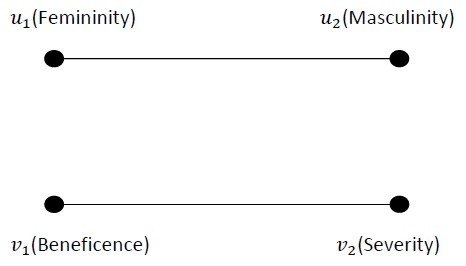}
      	\caption{A primary graph that represents a certain kind of stereotype. Labels are represented by black circles as vertices. $u_{1}$, $u_{2}$ respectively refer to a pair of labels: ``Femininity'' and ``Masculinity''. Meanwhile, $v_{1}$, $v_{2}$ respectively refer to the other pair of labels: ``Beneficence'' and ``Severity''. Black edges indicate that two adjacent vertices (labels) are complementary.}
      	\label{Figure 1}
      \end{figure}

      We use a vertex to represent each label separately. In addition, we assume that an edge exists between two vertices if and only if the corresponding labels are contradictory. According to the above descriptions, Figure \ref{Figure 1} shows a primary graph which indicates that ``Beneficence'' and ``Severity'', as well as ``Femininity'' and ``Masculinity'', are complementary.

      The relation between these two pairs of labels is not given yet, which is related to another kind of stereotypical thought. Suppose Anne doesn't believe that a woman can leave others an impression of severity. That is to say, ``People who are feminine can never be severe.'' There is also a reverse description:``People who are severe can never be feminine.'' Both descriptions lead to the same conclusion: People who are both feminine and severe don't exist. It conforms to the fact that many people may think it impossible to link femininity with severity. From this perspective, ``Femininity'' and ``Severity'' are contradictory.

\begin{figure}
      	\centering
      	\includegraphics[height=4cm]{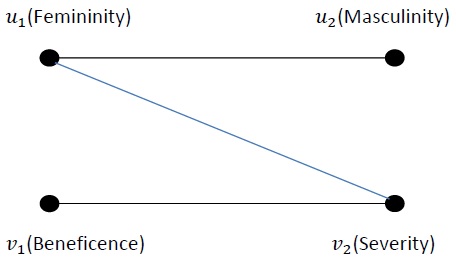}
      	\caption{A new graph based on Figure \ref{Figure 1} by adding $u_{1}v_{2}$. The new edge is colored in blue, indicating that ``Femininity''($u_{1}$) and ``Severity''($v_{2}$) are just contradictory.}
      	\label{Figure 2}
      \end{figure}

      In the graph given in Figure \ref{Figure 1}, we may add an edge which joins $u_{1}$ and $v_{2}$, denoted as $u_{1}v_{2}$. Hence we obtain a new graph shown in Figure \ref{Figure 2}, in which ``Femininity'' and ``Severity'' are just contradictory. Note that here we don't consider ``Femininity'' and ``Severity'' as a \emph{pair} of labels. Neither do we consider ``Masculinity'' and ``Beneficence''.
      
      We use set theory to describe the relation between ``Femininity'' and ``Beneficence''. Define following nonempty sets
      \begin{center}
        $\mathcal{F} =$ \{People who are feminine\},

        $\mathcal{M} =$ \{People who are masculine\},

        $\mathcal{B} =$ \{People who are beneficent\},

        $\mathcal{S} =$ \{People who are severe\} and

        $\mathcal{U} =$ \{People\}.
      \end{center}

      Let $E(G)$ and $V(G)$ denote the set of vertices and edges of the graph $G$ respectively.

      Given $G$ in Figure \ref{Figure 2}, we have
      \begin{equation*}
        u_{1} v_{2} \in E(G) \Rightarrow \mathcal{F} \cap \mathcal{S}=\emptyset.
      \end{equation*}

      Meanwhile, since ``Beneficence'' and ``Severity'' are complementary,
      \begin{equation*}
        \mathcal{B}= \mathcal{U}-\mathcal{S}.
      \end{equation*}

      Thus we have
      \begin{equation*}
        \label{Formula1}
        \mathcal{F}\subseteq \mathcal{B},\tag{1}
      \end{equation*}

      which means every feminine person is beneficent. It seems plausible, yet it is just a stereotypical thought.

      Similarly, we also have
      \begin{equation*}
        \label{Formula2}
        \mathcal{S}\subseteq \mathcal{M},\tag{2}
      \end{equation*}

      which means every severe person is masculine.

      The discussion above via set theory is an example of the relation between two labels which belong to different pairs. It is decided by whether or not an edge exists between two corresponding vertices. An interesting question is how many edges (i.e. new stereotypical thoughts) we can add to the primary graph in Figure \ref{Figure 1} without inducing paradoxes. More edges indicate more assumptions about relations among labels, which may lead to a paradox. We wonder if the paradox could occur when we obtain two (or even more) new edges.

\begin{figure}
      	\centering
      	\includegraphics[height=4cm]{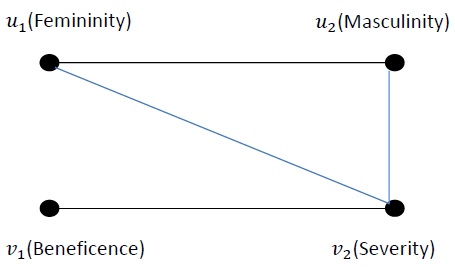}
      	\caption{A new graph based on Figure \ref{Figure 1} by adding $u_{1}v_{2}$ and $u_{2}v_{2}$. It includes a cycle $u_{1}v_{2}u_{2}$ that reflects a logical error.}
      	\label{Figure 3}
      \end{figure}

      New edges added to Figure \ref{Figure 1} will be denoted by blue lines in the figures, which indicate that the labels are just contradictory. First, we assume that two new edges share one vertex which belongs to one pair. Then they have the other different vertices which belong to the same pair. An example is given in Figure \ref{Figure 3}. Based on the stereotype in Figure \ref{Figure 1}, we obtain two new stereotypical thoughts: ``Femininity'' and ``Severity'' are contradictory, so are ``Masculinity'' and ``Severity''.

      It can easily show that a paradox is included. From the path $v_{1}v_{2}u_{1}u_{2}$, we have $\mathcal{S} \subseteq \mathcal{M}$, which is Formula (\ref{Formula2}). On the other hand, from the path $v_{1}v_{2}u_{2}u_{1}$, we have $\mathcal{M} \subseteq \mathcal{B}$. Thus $\mathcal{S}\subseteq \mathcal{M} \subseteq \mathcal{B}$, and $\mathcal{S}=\mathcal{B} \cap \mathcal{S} = \emptyset$, which contradicts the assumption ``for every label, there always exist someone who can be categorized to it''. It represents a kind of paradox that we called \emph{logical error}\footnote{The logical error indicates that the linkage we built among labels is not transitive. In Figure \ref{Figure 3}, an $u_{1}u_{2}$ edge and $u_{1}v_{2}$ edge lead to the irrational existence of an $u_{2}v_{2}$ edge, which represents the so-called logical error. Thus we may not regard linkage as an equivalence relation. Relations among characteristics in this article are nothing more than contradictory, complete, or complementary labels.}.

      In general, the \emph{logical error} in the stereotype is a self-contradictory conclusion derived from reasoning only based on the stereotype itself. It is quite fatal, since it indicates the present stereotype is nonsense in a logical sense. Once such of logical error is detected, the modification is required. In another word, a stereotype with logical errors is hard to linger in the cognition without change, which makes it unstable.

\begin{figure}
      	\centering
        \includegraphics[height=4cm]{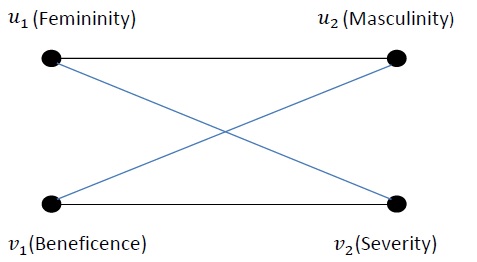}
        \caption{The first basic stereotype induced graph based on Figure \ref{Figure 1}. It is isomorphic to a cycle of length 4.}
        \label{Figure 4}
      	\centering
        \includegraphics[height=4cm]{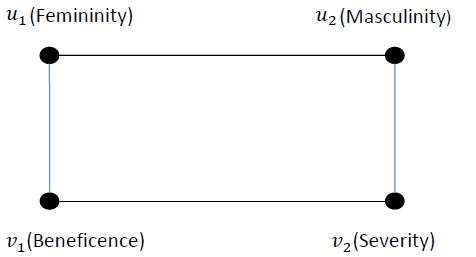}
        \caption{The second basic stereotype induced graph based on Figure \ref{Figure 1}. It is isomorphic to $C_{4}$ as Figure \ref{Figure 4}, although Figures \ref{Figure 4} and \ref{Figure 5} represent different stereotypical thoughts. }
        \label{Figure 5}
      \end{figure}

      To avoid logical errors, new edges added to Figure \ref{Figure 1} can never share the same vertex, which means there are at most two new edges. With two new edges added, there are just two different possible graphs based on Figure \ref{Figure 1}. Specific graphs are given in Figures \ref{Figure 4} and \ref{Figure 5}. We call them \emph{basic stereotype induced graphs}.

      One crucial point is that Figures \ref{Figure 4} and \ref{Figure 5} are both isomorphic to a cycle of length 4 (or quadrilateral). Let $C_{n}$ denote a cycle of length $n$, which is called $n$-cycle, where $n\ge2$ and $n\in\mathbb{N}$. According to the discussion above, we have
      \begin{description}
	    \item[Proposition 1]  A graph is a basic stereotype induced graph if and only if it is isomorphic to $C_{4}$.
        \item[Proposition 2]  A graph is a basic stereotype induced graph if and only if it is isomorphic to $K_{2,2}$\footnote{$K_{n,m}$ is the complete bipartite graph, which is constructed by: Giving $n$ vertices for the first set and $m$ vertices for the second set, then joining every vertex of the first set to every vertex of the second set.}.
      \end{description}
      \begin{proof}
        This is because $C_{4}$ is isomorphic to $K_{2,2}$.

        \qedhere
      \end{proof}

      We have a short discussion on how basic stereotype induced graphs avoid logical errors. Consider the graph $G$ refers to Figure \ref{Figure 4}. From the path $v_{1}v_{2}u_{1}u_{2}$, we have $\mathcal{S} \subseteq \mathcal{M}$, and from the path $v_{2}v_{1}u_{2}u_{1}$, we have $\mathcal{M} \subseteq \mathcal{S}$. Then $\mathcal{S}=\mathcal{M}$, which means ``Severity'' and ``Masculinity'' are identical. Similarly, $\mathcal{F}=\mathcal{B}$, which means ``Beneficence'' and ``Femininity'' are identical. These two conclusions are just prejudices, but not paradoxes, as they don't have any logical mistakes and even seem plausible. Thus, we can say the basic stereotype induced subgraph holds no logical errors, and vice versa.

      Given a family of sets $\mathcal{A}=\{X_{1},X_{2},\dots,X_{n}\}$ where $X_{i}$ $(i\in\{1,2,\dots,n\})$ is the subset of the vertex set of graph $\Gamma$. The \emph{intersection graph} of $\mathcal{A}$, denoted $\Omega(\mathcal{A})$, is the graph having $\mathcal{A}$ as vertex set with $X_{i}$ adjacent to $X_{j}$ $(i\ne j)$ if and only if $\exists u\in X_{i},\exists v\in X_{j}, s.t.$ $uv \in E(\Gamma)$ \citep{McKee1999}.

\begin{figure}
      	\centering
      	\includegraphics[height=1.5cm]{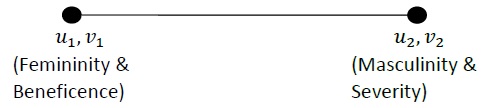}
      	\caption{The intersection graph based on Figure \ref{Figure 4}. ``Femininity'' and ``Beneficence'' are merged to the left clique while ``Masculinity'' and ``Severity'' are merged to the right clique. That means in one's stereotype, the notions of ``Severity'' and ``Masculinity'' are identical, as well as ``Beneficence'' and ``Femininity''. We use black circles to represent cliques rather than individual labels in the intersection graph, since Labels in the same clique imply the identity of characteristics.}
      	\label{Figure 6}
      	\centering
      	\includegraphics[height=1.5cm]{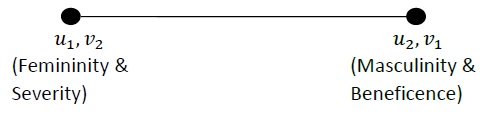}
      	\caption{The intersection graph based on Figure \ref{Figure 5}. ``Femininity'' and ``Severity'' are merged to the left clique while ``Masculinity'' and ``Beneficence'' are merged to the right clique, which makes Figure \ref{Figure 5} represent different stereotypical thoughts from Figure \ref{Figure 4}.}
      	\label{Figure 7}
      \end{figure}

      In Figure \ref{Figure 4}, let $X_{1}=\{u_{1},v_{1}\}$ and $X_{2}=\{u_{2},v_{2}\}$. Each set contains two labels that are not adjacent before. Actually, $X_{1}$ and $X_{2}$ can be regarded as equivalence classes, consisting of identical labels respectively. Then $\Omega(\{X_{1},X_{2}\})$ is shown in Figure \ref{Figure 6}. Thus, we can easily understand that in this graph, vertices which are not adjacent before become identical. It indicates that in one's stereotype, the notions of ``Severity'' and ``Masculinity'' are identical, as well as ``Beneficence'' and ``Femininity''.

      The transportation from Figure \ref{Figure 4} to Figure \ref{Figure 6} is considered a reasonable speculation in one's cognition. It seems that initially different labels are merged into cliques, and two pairs of complementary labels become one pair. We intuitively call it ``merge operation'' until we give the mathematical definition of such a crucial process in the next section (see Definition 5).

      Since the merge operation recognizes identical labels and merges them (which seems somewhat tedious if we still regard them as different labels), one of its benefits is to simplify the stereotype. The stereotype itself serves an individual function by simplifying the perceived information \citep{Tajfel1981}, thus we believe the merge operation is reasonable when the tedious categorization occurs.

      A merge operation on Figure \ref{Figure 5} leads to Figure \ref{Figure 7}. While it doesn't work on some graphs like Figure \ref{Figure 3}, as no two labels can be regarded as identical. The graph of Figure \ref{Figure 3} is not a basic stereotype induced graph. On the other hand, a merge operation acting on a basic stereotype graph will induce $K_{2}$\footnote{$K_{n}$ is the complete graph, which is constructed by: Giving $n$ vertices and then joining every vertex to every other. In this way, $K_{2}$ looks just like a line.}.

    \subsection{Mathematical Definition of Stereotype Graphs}
      We no longer list specific labels afterwards unless special descriptions occur. Instead, we will give the mathematical definition of stereotype graphs in this subsection.

      We expand complementary labels to $n$ pairs $(n\in \mathbb{Z}_{+})$. As for the graph, we use vertices $u_{1}^{i}$ and $u_{2}^{i}$ to denote the $i$th pair of labels respectively $(i\in \{1,2,\dots,n\})$. In addition, let $\Gamma_{n}$ denote the stereotype graph which has $n$ pairs of labels.

      Note that every two pairs of labels can have a similar relation as in Figures \ref{Figure 4} and \ref{Figure 5}. Based on the discussions above, we give the mathematical definition of \emph{stereotype graph} below.
      \begin{description}
	    \item[Definition 3] For $n\in \mathbb{Z}_{+}$, a graph that satisfies all of the following conditions is called the \emph{stereotype graph $\Gamma_{n}$}:
        \begin{enumerate}
          \item It has $n$ pairs of vertices, denoted by $u_{1}^{i}$ and $u_{2}^{i}$ $(i\in \{1,2,\dots,n\})$. For each $i$, $u_{1}^{i}$ and $u_{2}^{i}$ are a \emph{pair} of vertices with \emph{pair index} $i$;
          \item $\forall i\in [1,n]$, $u_{1}^{i}u_{2}^{i}\in E(\Gamma_{n})$;
          \item For any $i, j\in \mathbb{Z}_{+}, 1\le i<j\le n$, the induced subgraph $u_{1}^{i}u_{2}^{i}u_{1}^{j}u_{2}^{j}$ is a basic stereotype induced subgraph (i.e. isomorphic to $K_{2,2}$).
        \end{enumerate}
        \item[Notation 4]  $ST_{n}$ is the set of all stereotype graphs $\Gamma_{n}$ with $n$ pairs of vertices $(n\in \mathbb{Z}_{+})$.
      \end{description}

      Based on the stereotype graph, we can directly give a mathematical definition of the so-called \emph{merge operation} mentioned in the previous section.
      \begin{description}
	    \item[Definition 5] Suppose there is a stereotype graph $\Gamma$ with $m$ vertices, denoted $x_{1},x_{2},\dots,x_{m}$ $(m\ge4)$. For any induced subgraph $x_{i}x_{j}x_{k}x_{l}$ ($i, j, k$ and $l$ are pairwise unequal, $i<k$) of $\Gamma$, if $x_{i}x_{j}x_{k}x_{l}$ is a basic stereotype induced graph with $x_{i}x_{j}, x_{k}x_{l}\notin E(\Gamma)$, let
        \begin{equation*}
            X_{t}=
            \left\{
            \begin{array}{rl}
                \{x_{t}\} \quad & (t\notin\{i,j,k,l\}) \\
                \{x_{i},x_{j}\} \quad & (t=j) \\
                \{x_{k},x_{l}\} \quad & (t=l) \\
            \end{array}
        \right.
        \end{equation*}
        and $\mathcal{A}=\{X_{1},\dots,X_{i-1},X_{i+1},\dots,X_{k-1},X_{k+1},\dots,X_{n}\}$. The transformation from $\Gamma$ to $\Omega(\mathcal{A})$ is called the \emph{merge operation} (acting) on $x_{i}x_{j}x_{k}x_{l}$.
      \end{description}

\begin{figure}
      	\centering
        \includegraphics[height=4cm]{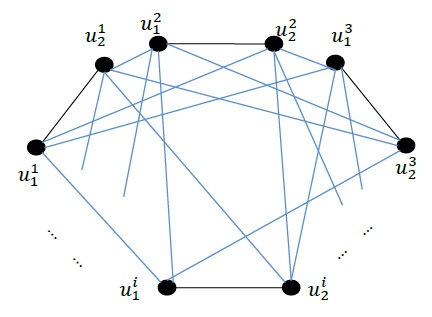}
        \caption{An example of a stereotype graph with labels represented by vertices (black circle). Black edges refer to complementary relations, while blue edges refer to contradictory relations.}
        \label{Figure 8}
      \end{figure}

      To provide an intuitive illustration of the stereotype graph, a general stereotype graph is shown in Figure \ref{Figure 8}.

      We then make several statements about the correspondence between the general stereotype graphs defined by Definition 3 and the category stereotype. These statements link the abstract stereotype graph and the psychological category stereotype.
      \begin{enumerate}
        \item Vertices and edges have their psychological meanings. \emph{Vertices} represent \emph{labels}, or \emph{characteristics}, in the category stereotype. Meanwhile, two adjacent vertices (i.e. \emph{edges} in Definition 3(2)) indicate that their corresponding labels are \emph{contradictory}. Relations among edges are not transitive.
        \item Vertices and edges construct a \emph{simple graph} which portrays one's cognitive thinking that may be described as a stereotype. The stereotype here refers to the category stereotype, which is a personal cognitive belief about judging people as category members rather than individuals. \citep{Perry2000a}
        \item In Definition 3(1), we consider $n$ \emph{pairs} of \emph{complementary labels} in one's stereotype. For two labels within the same pair, the adjacency between two related vertices indicates \emph{complementariness} but not just contradictoriness.
        \item  We assume that for every label, there always exists someone who can be categorized under it. This ensures labels are not meaningless in the stereotype graph.
        \item Adding a new edge to the graph means accepting the corresponding stereotypical thought. Deleting an edge from the graph means abandoning the corresponding stereotypical thought.
        \item We use \emph{merge operation} (Definition 5) to recognize and merge identical labels into cliques so that we can simplify the stereotype and facilitate the stereotyping process.
        \item The \emph{logical error} in the stereotype is a self-contradictory conclusion derived from reasoning only based on the stereotype itself. It is reasonable that if there's a logical error in a stereotype, the stereotype needs to be modified and thus be harder to stay in the cognition. In another word, the stereotype with logical errors is \emph{unstable} while the stereotype without logical errors is \emph{stable}.
        \item Adding a new edge may increase the risk of producing logical errors while removing an edge may reduce the risk. Moreover, we may detect the logical error through merge operations, as we will explain in a later section.
        \item Definition 3(3) ensures that the stereotype graph defined is relatively the most unstable one. This is because we will detect logical errors whenever we add any new edge (i.e. new stereotypical thought).
        \item We still need to show whether the stereotype portrayed by Definition 3 always exists or not. The bipartite criterion (Theorem 17) in the later section ensures the existence.
      \end{enumerate}

      Henceforward, we don't sedulously mention the psychological meaning of stereotype graphs unless necessary.

      It's worth mentioning Definition 3(3) again. It seems so strict that it is hard to exist. The function of Definition 3(3) is to consider the relatively extreme circumstance (here refers to the relatively most unstable stereotype, as mentioned in the ninth statement), as well as to simplify the model of stereotype graphs. If we remove Definition 3(3), we may greatly enlarge the applying scope of stereotype graphs. For example, let's consider the circumstance that someone categorizes people according to their nationalities (e.g. American, English, French, etc.). Although the original categorization can't be portrayed by pairs of complementary labels, we can add some extra labels to make them dually complementary (e.g. American \& non-American, English \& non-English, etc.). However, we will lose the property described by Definition 3(3). We mention circumstances without Definition 3(3) in the discussion section.

      Several straightforward properties of the stereotype graph are given below.
      \begin{description}
	    \item[Proposition 6]  $\forall\Gamma_{n}\in ST_{n}$ $(n\in \mathbb{Z}_{+})$,
        \begin{enumerate}
          \item $|V(\Gamma_{n})|=2n$, $|E(\Gamma_{n})|=2\cdot\binom{n}{2}+n=n^2$;
          \item $\Gamma_{n}$ is $n$-regular\footnote{A graph is $n$-regular means any vertex's degree is $n$}.
          \item The girth\footnote{Girth is the length of one of the shortest cycles as an induced subgraph of the graph.} of $\Gamma_{n}$ is no more than 4, i.e. the girth of $\Gamma_{n}$ $(n\ge 2)$ is 3 or 4;
          \item The diameter (i.e. the greatest distance between any two vertices) of $\Gamma_{n}$ is 2 except $\Gamma_{1}$;
          \item $\Gamma_{n}$ is connected;
        \end{enumerate}
      \end{description}

    \subsection{Bipartite Stability and Bipartite Criterion}
      In this section, we give a mathematical approach to portrait how stable can a stereotype stay on one's cognition. Several criteria can be used to decide whether a stereotype can avoid logical errors or not. The first one, called the bipartite criterion, is based on the merge operation.

\begin{figure}
      	\centering
        \includegraphics[height=4cm]{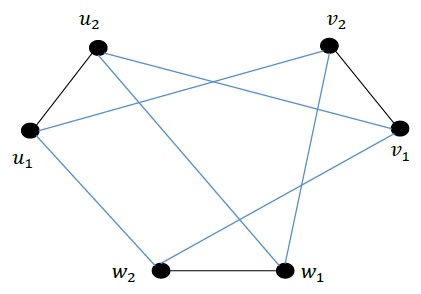}
        \caption{The first type of stereotype graph in $ST_{3}$, denoted $\Gamma_{3}^{(1)}$. Three pairs of labels are represented by vertices (black circles). Black edges refer to complementary relations, while blue edges refer to contradictory relations.}
        \label{Figure 9}
        \centering
        \includegraphics[height=4cm]{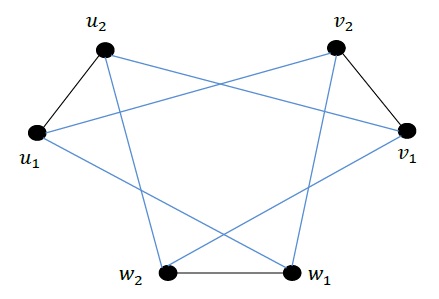}
        \caption{The second type of stereotype graph in $ST_{3}$, denoted $\Gamma_{3}^{(2)}$. How Figures \ref{Figure 9} and \ref{Figure 10} represent different stereotypical thoughts is explained in Figures \ref{Figure 11} and \ref{Figure 12}.}
        \label{Figure 10}
      \end{figure}

      First we consider a simple example. Suppose there are 3 pairs of labels, denoted $(u_{1},u_{2}),(v_{1},v_{2})$ and $(w_{1},w_{2})$. The stereotype graph $\Gamma_{3}$ is not unique. Two of the graphs in $ST_{3}$ are shown in Figures \ref{Figure 9} and \ref{Figure 10}.

\begin{figure}
      	\centering
      	\includegraphics[height=4cm]{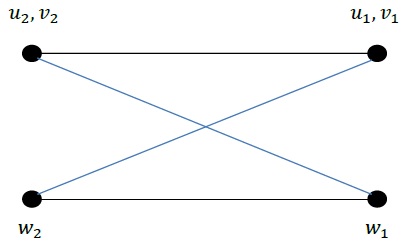}
      	\caption{The intersection graph $\Omega^{(1)}(\{X_{1},X_{2}\})$ induced by the merge operation acting on $u_{1}u_{2}v_{1}v_{2}$ in $\Gamma_{3}^{(1)}$. As the graph is a basic stereotype-induced graph, it can be acted on by another merge operation. The result is $K_{2}$, which is given in Figure \ref{Figure 13}.}
      	\label{Figure 11}
      	
      	\centering
      	\includegraphics[height=4cm]{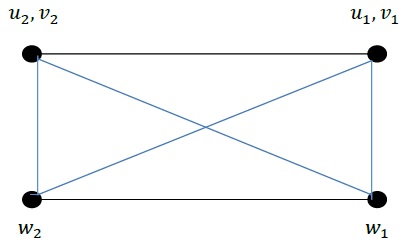}
      	\caption{The intersection graph $\Omega^{(2)}(\{X_{1},X_{2}\})$ induced by the merge operation acting on $u_{1}u_{2}v_{1}v_{2}$ in $\Gamma_{3}^{(2)}$. The merge operation can't work on it because of 3-cycles as an induced subgraph, indicating the existence of a logical error.}
      	\label{Figure 12}
      \end{figure}

      Now we do a merge operation (see Definition 5) acting on the induced subgraph $u_{1}u_{2}v_{1}v_{2}$. Let $X_{1}=\{u_{2},v_{2}\}$ and $X_{2}=\{u_{1},v_{1}\}$ . Then the two types of graphs above will turn to two intersection graphs shown in Figures \ref{Figure 11} and \ref{Figure 12}.

\begin{figure}
      	\centering
      	\includegraphics[height=1cm]{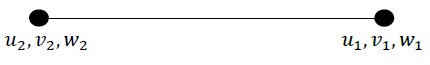}
      	\caption{By doing two merge operations (From Figure \ref{Figure 9} to Figure \ref{Figure 11}, and from Figure \ref{Figure 11} to Figure \ref{Figure 13}), $\Gamma_{3}^{(1)}$ is turned to $K_{2}$. In this way, $u_{2}$, $v_{2}$, and $w_{2}$ are considered as identical labels. Meanwhile, the other three labels are merged to the right clique, holding complementary relations with $u_{2}$, $v_{2}$, and $w_{2}$.}
      	\label{Figure 13}
      \end{figure}

      $\Omega^{(1)}(\{X_{1},X_{2}\})$ (Figure \ref{Figure 11}) is a basic stereotype-induced graph, thus it can be acted on by another merge operation. The result is $K_{2}$, which is given in Figure \ref{Figure 13}. We also have the conclusion from Figure \ref{Figure 13} that $u_{2},v_{2}$ and $w_{2}$ are identical labels, while $u_{1},v_{1}$ and $w_{1}$ are identical.

      However, the merge operation can't work on $\Omega^{(2)}(\{X_{1},X_{2}\})$ (Figure \ref{Figure 12}) since every two vertices are adjacent. Actually, there are two blue edges sharing the same vertex, which indicates a logical error according to the Figure \ref{Figure 3}.

      Such process is conducted after a merge operation acting on the induced subgraph $u_{1}u_{2}v_{1}v_{2}$. When we do the merge operation acting on the induced subgraph $u_{1}u_{2}w_{1}w_{2}$ or $v_{1}v_{2}w_{1}w_{2}$, we also find that $\Gamma_{3}^{(1)}$ can turn to $K_{2}$ by merge operations while $\Gamma_{3}^{(2)}$ can't. The reason that $\Gamma_{3}^{(2)}$ can't turn to $K_{2}$ is because there will emerge a logical error when doing merge operations, which is contrary to $\Gamma_{3}^{(1)}$. Thus we can say that in comparison to $\Gamma_{3}^{(2)}$, $\Gamma_{3}^{(1)}$ is more stable in one's cognition.

      Now we consider the generalized condition. We give a stereotype graph $\Gamma_{n}\in ST_{n}$ $(n\in \mathbb{Z}_{+})$. Since a merge operation reduces the number of vertices by 2, there will be $n-1$ steps of merge operations to turn $\Gamma_{n}$ in, if possible, $K_{2}$. Thus, all graphs in $ST_{n}$ can be divided into two parts: Those which can turn to $K_{2}$ after $n-1$ steps of merge operations, and those which can't. A graph in the part avoids logical errors, so it is more stable. A graph in the latter part avoids type contains logical errors, so it is more unstable.
      \begin{description}
	    \item[Definition 7]  $\forall\Gamma_{n}\in ST_{n}$ $(n\in \mathbb{Z}_{+})$, $\Gamma_{n}$ is \emph{bipartitely stable} if $\Gamma_{n}$ can be turned in $K_{2}$ through $n-1$ steps of merge operations; $\Gamma_{n}$ is \emph{bipartitely unstable} otherwise. The property of whether $\Gamma_{n}$ is bipartitely stable or bipartitely unstable is called \emph{bipartite stability}.
        \item[Notation 8]  Set $STB_{n}=\{\Gamma_{n}|\Gamma_{n}\in ST_{n}$ and $\Gamma_{n}$ is bipartitely stable\}.
      \end{description}

      The word bipartite derives from $K_{2}$, where $n$ pairs of labels are divided into two equivalence classes. In fact, as a terminology of graph theory, the word bipartite has a similar meaning. If you already have basic knowledge about graph theory, you may comprehend it easily when you read the next section.

      Consider the sequence of merge operations. When two steps of merge operations exchange their contiguous position in the operation sequence with each other, will the intersection graph change after these two operations? If won't change, the arbitrary change of the sequence of merge operations would never affect the final result.
      \begin{description}
	    \item[Notation 9]  The graph after a merge operation which acts on the basic stereotype induced subgraph $u_{1}u_{2}v_{1}v_{2}$ of a graph $\Gamma$ is denoted by $\Omega(\Gamma,u_{1}u_{2}v_{1}v_{2})$.
      \end{description}

      To simplify the description, we need to modify the representation of vertices in intersection graphs. Given $\Gamma_{n}$ with $n$ pairs of vertices $(u_{1}^{i},u_{2}^{i})$, we will have two equivalence classes as two new vertices of the intersection graph after a merge operation. At this time, we use $u_{1}^{i}$ and $u_{2}^{i}$ to denote those two vertices, where $i$ is the minimum number of the pair index in equivalence classes. Then the intersection graph can be deemed again a stereotype graph, waiting for the next merge operation.
      \begin{description}
	    \item[Theorem 10]  $\forall n\in\mathbb{Z}_{+}$ $(n\ge3), \forall \Gamma\in ST_{n}$,
        for arbitrary three different pairs of vertices $(u_{1}^{i},u_{2}^{i}),(u_{1}^{j},u_{2}^{j})$ and $(u_{1}^{k},u_{2}^{k})$ $(i<j<k)$, we have, if all merge operations are possible,
          \begin{equation*}
             \Omega(\Omega(\Gamma,u_{1}^{i}u_{2}^{i}u_{1}^{j}u_{2}^{j}),u_{1}^{i}u_{2}^{i}u_{1}^{k}u_{2}^{k})
          \end{equation*}
          is the same as
          \begin{equation*}
             \Omega(\Omega(\Gamma,u_{1}^{i}u_{2}^{i}u_{1}^{k}u_{2}^{k}),u_{1}^{i}u_{2}^{i}u_{1}^{j}u_{2}^{j})
          \end{equation*}
      \end{description}

      \begin{proof}
        Suppose in $\Omega(\Omega(\Gamma,u_{1}^{i}u_{2}^{i}u_{1}^{j}u_{2}^{j}),u_{1}^{i}u_{2}^{i}u_{1}^{k}u_{2}^{k})$, $u_{p}^{i}, u_{q}^{j}$ and $u_{r}^{k}$ are in the same equivalence class, while $u_{3-p}^{i},u_{3-q}^{j}$ and $u_{3-r}^{k}$ are in the other equivalence class.

        Then in the graph $\Gamma, u_{p}^{i},u_{q}^{j}$ and $u_{r}^{k}$ are not adjacent to each other, as well as $u_{3-p}^{i},u_{3-q}^{j}$ and $u_{3-r}^{k}$ are not adjacent to each other either.

        Thus in $\Omega(\Omega(\Gamma,u_{1}^{i}u_{2}^{i}u_{1}^{k}u_{2}^{k}),u_{1}^{i}u_{2}^{i}u_{1}^{j}u_{2}^{j}), u_{p}^{i} ,u_{q}^{j}$ and $u_{r}^{k}$ are in the same equivalence class, while $u_{3-p}^{i}, u_{3-q}^{j}$ and $u_{3-r}^{k}$ are in the other equivalence class. This is the same as the result in the graph $\Omega(\Omega(\Gamma,u_{1}^{i}u_{2}^{i}u_{1}^{j}u_{2}^{j}),u_{1}^{i}u_{2}^{i}u_{1}^{k}u_{2}^{k})$.

        In addition, the process won't affect vertices $(u_{1}^{l}, u_{2}^{l})$ for any $l\ne i,j,k$.

        Thus, the two graphs are the same.

        \qedhere
      \end{proof}

      \begin{description}
	    \item[Theorem 11]  $\forall n\in\mathbb{Z}_{+}$ $(n\ge4)$, $\forall \Gamma\in ST_{n}$, for arbitrary four different pairs of vertices $(u_{1}^{i},u_{2}^{i}),(u_{1}^{j},u_{2}^{j}),(u_{1}^{k},u_{2}^{k})$ and $(u_{1}^{l},u_{2}^{l})$ $(i<j, k<l)$, we have, if all merge operations are possible,
          \begin{equation*}
             \Omega(\Omega(\Gamma,u_{1}^{i}u_{2}^{i}u_{1}^{j}u_{2}^{j}),u_{1}^{k}u_{2}^{k}u_{1}^{l}u_{2}^{l})
           \end{equation*}
           is the same as
           \begin{equation*}
             \Omega(\Omega(\Gamma,u_{1}^{k}u_{2}^{k}u_{1}^{l}u_{2}^{l}),u_{1}^{i}u_{2}^{i}u_{1}^{j}u_{2}^{j})
          \end{equation*}
      \end{description}

      \begin{proof}
        The proof procedure is similar to the proof of Theorem 10.

        \qedhere
      \end{proof}

      \begin{description}
	    \item[Corollary 12]  $\forall n\in\mathbb{Z}_{+}, \forall\Gamma_{n}\in ST_{n}$, if we do merge operations on it arbitrarily and repeatedly until no more merge operations can be done, the final graphs we have won't change when previous merge operations change.
      \end{description}

      \begin{proof}
        We can easily prove this by using Theorem 10, Theorem 11, and exchanging contiguous merge operations repeatedly.

        \qedhere
      \end{proof}

      Now it's time to decide how to judge a stereotype graph $\Gamma_{n}$ is bipartitely stable or not. Consider that we are going to color the vertices. Each vertex just holds one color. The only restriction is no adjacent vertices can hold the same color. In graph theory, it is called \emph{vertex coloring}.

      $\forall\Gamma_{n}\in STB_{n}$, since it can turn to $K_{2}$, the vertices of $\Gamma_{n}$ can be divided into two parts (which are actually equivalence classes). Vertices in the same part are not adjacent to each other. Thus, we can use just two colors to color $\Gamma_{n}$, with each color indicates an equivalence class (consisting of labels which are actually the same with each other). However, $\forall\Gamma_{n}\notin STB_{n}$, two colors are not enough since vertices can't be divided into two equivalence classes.
  
      \begin{description}
	    \item[Definition 13 \citep{Jensen2011}]  Suppose there is a graph $\Gamma$ with $n$ vertices, denoted $v_{1},v_{2},\dots,v_{n}$. $\Gamma$ is \emph{$k$-colorable} if there exists a mapping
\begin{center}
$\theta:V(\Gamma)=\{v_{1},v_{2},\dots,v_{n}\}\rightarrow\{1,2,\dots,k\}$
\end{center}
where $\theta(v_{i})\ne \theta(v_{j})$ for any $v_{i}v_{j}\in E(\Gamma)$. Such a mapping $\theta$ is called a \emph{vertex $k$-coloring}. If $\theta$ is a surjection, it is called a \emph{proper vertex $k$-coloring}.
      \end{description}

      In a more comprehensive way to say, in vertex coloring, a graph is $k$-colorable if we can use $k$ colors to color it, with no adjacent vertices own the same color.
      \begin{description}
	    \item[Definition 14 \citep{Biggs1974P52}]  Suppose there is a graph $\Gamma$ with $n$ vertices, denoted $v_{1},v_{2},\dots,v_{n}$. A \emph{vertex coloring $i$-partition} of $\Gamma$ is a set partition $V(\Gamma)=\bigcup_{j=1}^{i}{V_{j}}$  where $V_{j}\ne\emptyset$ and no two adjacent vertices are in the same set.
      \end{description}

      \begin{description}
	    \item[Theorem 15 (Coloring Criterion)]  $\forall\Gamma_{n}\in ST_{n}$, we have
        \begin{center}
          $\Gamma_{n}\in STB_{n}\Leftrightarrow \Gamma_{n}$ is 2-colorable.
        \end{center}
      \end{description}

      The coloring criterion has a unique psychological meaning. According to the equivalent class mentioned before, vertices with the same color indicate that the corresponding labels are identical. Thus, we say a stereotype is stable (i.e. contains no logical errors) only when there're only two different kinds of labels in the stereotype. This resembles a traditional idea that stereotyping arose out of the processes of ``ordinary'' cognition which was neither ``faulty'' nor ``correct'' \citep{Perry2000a}.

      The following proposition is a basic result in graph theory.
      \begin{description}
	    \item[Proposition 16 \citep{Konig1916}]  For a graph $\Gamma$, the following three statements are equivalent.
        \begin{enumerate}
          \item $\Gamma$ is 2-colorable;
          \item $\Gamma$ contains no odd cycles (the odd cycle is a cycle whose length is odd);
          \item $\Gamma$ is a bipartite graph.
        \end{enumerate}
      \end{description}

      Finally, according to the discussion above, we give an important criterion for judging bipartite stability.
      \begin{description}
	    \item[Theorem 17 (Bipartite Criterion)]  $\forall\Gamma_{n}\in ST_{n}$, we have
          \begin{center}
            $\Gamma_{n}\in STB_{n}\Leftrightarrow \Gamma_{n}$ is isomorphic to $K_{n,n}$.
          \end{center}
      \end{description}

      \begin{proof}
        Since Theorem 15,
        \begin{equation*}
          \label{Formula3}
          \Gamma_{n}\in STB_{n}\Leftrightarrow\Gamma_{n}\mbox{ is 2-colorable.} \tag{3}
        \end{equation*}

        Since Proposition 16,
        \begin{equation*}
          \label{Formula4}
          \Gamma_{n}\mbox{ is 2-colorable}\Leftrightarrow \Gamma_{n} \mbox{ is a bipartite graph}. \tag{4}
        \end{equation*}

        If $\Gamma_{n}$ is isomorphic to $K_{n,n}$, $\Gamma_{n}$ is actually a bipartite graph.

        Conversely, if $\Gamma_{n}$ is a bipartite graph, since $\Gamma_{n}\in ST_{n}$, according to Definition 3, we have
        \begin{equation*}
          u_{1}^{i}u_{2}^{i}\in E(\Gamma_{n}), \forall i\in\{1,2,\dots,n\}.
        \end{equation*}

        Thus given its partite sets $\mathcal{U}$ and $\mathcal{V}$, with $\mathcal{U}\cap\mathcal{V}=\emptyset$ and $\mathcal{U}\cup\mathcal{V}=V(\Gamma_{n})$, we have $|\mathcal{U}|=|\mathcal{V}|=n$.

        Since vertices in $\mathcal{U}$ are not adjacent to each other, and so as vertices in $\mathcal{V}$, according to Proposition 6(1), we have
        \begin{equation*}
          n^2=|E(\Gamma_{n})|\le|E(K_{n,n})|=|\mathcal{U}|\cdot|\mathcal{V}|=n^2.
        \end{equation*}

        Since $|E(\Gamma_{n})|=|E(K_{n,n})|$ if and only if $\Gamma_{n}$ is a complete bipartite graph, it indicates that $\Gamma_{n}$ is isomorphic to $K_{n,n}$.

        Thus,
        \begin{equation*}
          \label{Formula5}
          \Gamma_{n}\mbox{ is a bipartite graph}\Leftrightarrow\Gamma_{n}\mbox{ is isomorphic to }K_{n,n}.\tag{5}
        \end{equation*}

        From (3) to (5), we have
        \begin{equation*}
          \Gamma_{n}\in STB_{n}\Leftrightarrow \Gamma_{n}\mbox{ is isomorphic to }K_{n,n}.
        \end{equation*}

        \qedhere
      \end{proof}

      \begin{description}
	    \item[Corollary 18]  $STB_{1}=ST_{1}=\{K_{1}\},STB_{2}=ST_{2}=\{K_{2,2}\}.$
      \end{description}

      The bipartite criterion indicates that complete bipartite graphs are the same as bipartite stable stereotype graphs. It shows the structure of bipartitely stable stereotype graphs, which greatly facilitates us to judge the stability of a stereotype.

    \subsection{The Girth of Stereotype Graphs and Girth Criterion}
      Recall Proposition 6(3), we know that $\forall\Gamma_{n}\in ST_{n}$ $(n\ge2)$, the girth of $\Gamma_{n}$ can only be 3 or 4. Actually it is also related to bipartite stability.
      \begin{description}
	    \item[Notation 19]  The girth of a graph $\Gamma$ is denoted by $\mathcal{G}(\Gamma)$.
      \end{description}

      \begin{description}
	    \item[Theorem 20 (Girth Criterion)]  $\forall\Gamma_{n}\in ST_{n}$ $(n\ge2)$, we have
          \begin{equation*}
            \Gamma_{n}\in STB_{n}\Leftrightarrow\mathcal{G}(\Gamma_{n})=4.
          \end{equation*}
      \end{description}

      \begin{proof}
        If $\mathcal{G}(\Gamma_{n})=3$, then $\Gamma_{n}$ has a 3-cycle as an induced subgraph, where 3 vertices are adjacent to each other. It indicates that $\Gamma_{n}$ is not 2-colorable. By Theorem 15, we have $\Gamma_{n}\notin STB_{n}$.

        On the other hand, if $\Gamma_{n}\notin STB_{n},\Gamma_{n}$ can't turn to $K_{2}$ through any series of merge operations. Thus there exist three vertices which are adjacent to each other. Then $\mathcal{G}(\Gamma_{n})\le3$. Since $\mathcal{G}(\Gamma_{n})\ge3$ for any graph $\Gamma$, we have $\mathcal{G}(\Gamma_{n})=3$.

        Thus $\forall\Gamma_{n}\in ST_{n}, n\ge2$, we have
        \begin{equation*}
          \Gamma_{n}\notin STB_{n}\Leftrightarrow\mathcal{G}(\Gamma_{n})=3.
        \end{equation*}

        By Proposition 6(3), we know that $\mathcal{G}(\Gamma_{n})=3$ or 4, which means
        \begin{equation*}
          \Gamma_{n}\in STB_{n}\Leftrightarrow\mathcal{G}(\Gamma_{n})=4.
        \end{equation*}

        \qedhere
      \end{proof}

      \begin{description}
	    \item[Corollary 21]  $\forall\Gamma_{n}\in ST_{n}$ $(n\ge2)$, we have
          \begin{equation*}
            \Gamma_{n}\in STB_{n}\Leftrightarrow\Gamma_{n}\mbox{ contains no 3-cycles}.
          \end{equation*}
      \end{description}

      When a stereotype graph $\Gamma_{n}\notin STB_{n}$, as for any of the 3-cycle included in $\Gamma_{n}$, its three vertices come from different pairs of labels.

      The girth criterion shows a crucial property of bipartitely stable stereotype graphs without 3-cycles, which will play an important role in subsequent sections.

  \section{Algebraic Graph Theory in Stereotypes}
    Algebraic graph theory is deeply related to solving problems computationally. Therefore, we make use of algebraic graph theory to further discuss stereotype graphs.

    \subsection{The Adjacency Matrix of Stereotype Graphs}
      The following matrix, named the adjacency matrix, is the core of algebraic graph theory.
      \begin{description}
	    \item[Definition 22]  Suppose there is a graph $\Gamma$ with $n$ vertices, denoted $v_{1},v_{2},\dots,v_{n}$. For the matrix $A(\Gamma)=\{a_{i,j}\}_{n\times n}$, define $a_{i,j}=1$ if $v_{i}v_{j}\in E(\Gamma) ; a_{i,j}=0$ otherwise, where $1\le i,j\le n$. Such $A(\Gamma)$ is called the \emph{adjacency matrix} of $\Gamma$.
      \end{description}

      The $n\times n$ matrix whose entries are 1 is an important matrix and is denoted by $J_{n}$. A simple result we have is
      \begin{description}
	    \item[Proposition 23]  $\forall\Gamma_{n}\in ST_{n},A(\Gamma_{n})J_{2n}=J_{2n}A(\Gamma_{n})=nJ_{2n}.$
      \end{description}

      \begin{proof}
        This is because $\Gamma_{n}$ is $n$-regular.

        \qedhere
      \end{proof}

      When we run a program to analyse graphs, we actually input adjacency matrices or the equivalent adjacency lists. Thus if we want to show bipartite stability computationally, it would be better if the bipartite stability could directly emerge from adjacency matrices.

      \begin{description}
	    \item[Theorem 24]  $\forall\Gamma_{n}\in ST_{n},\Gamma_{n}\in STB_{n}$ if and only if $A(\Gamma_{n})$ has no principal minor $det(\begin{bmatrix}0&1&1\\1&0&1\\1&1&0\end{bmatrix})$. Here $det(M)$ is the determinant of the matrix $M$.
      \end{description}

      \begin{proof}
        The principal minor $det(\begin{bmatrix}0&1&1\\1&0&1\\1&1&0\end{bmatrix})$ actually indicates a 3-cycle. Thus we can prove the theorem by Corollary 21.

        \qedhere
      \end{proof}

      It requires effort to find out whether a certain kind of principal minor exists in an adjacency matrix. In comparison, the following theorem is more operational and convenient to use, since it just needs simple operations on matrices.
      \begin{description}
	    \item[Theorem 25 (Matrix Criterion)]  $\forall\Gamma_{n}\in ST_{n}$ $(n\ge2)$, we have
           \begin{equation*}
             \Gamma_{n}\in STB_{n}\Leftrightarrow A(\Gamma_{n})^2+nA(\Gamma_{n})=nJ_{2n}.
           \end{equation*}
      \end{description}

      The proof is complicated. We provide several lemmas before formally proving it.
      \begin{description}
	    \item[Definition 26 \citep{Bose1963}]  A \emph{strongly regular graph}, denoted as $srg(n,k,p,q)$, is a $k$-regular graph with $n$ vertices which satisfies
        \begin{enumerate}
          \item Every two adjacent vertices has $p$ common adjacent vertices;
          \item Every two non-adjacent vertices has $q$ common adjacent vertices.
        \end{enumerate}
      \end{description}

      \begin{description}
	    \item[Lemma 27]  $\forall\Gamma_{n}\in ST_{n},$ we have
        \begin{equation*}
          \Gamma_{n}\in STB_{n}\Leftrightarrow\Gamma_{n}\mbox{ is isomorphic to }srg(2n,n,0,n).
        \end{equation*}
      \end{description}

      \begin{proof}
        By Proposition 6, we already know that $\Gamma_{n}$ is an $n$-regular graph with $2n$ vertices.

        In terms of the girth criterion, $\Gamma_{n}\notin STB_{n}\Leftrightarrow\mathcal{G}(\Gamma_{n})=3$, i.e. it has 3-cycles. Thus, graphs not in $STB_{n}$ can't be isomorphic to $srg(2n,n,0,n)$ since there exist two adjacent vertices that share an adjacent vertex.

        On the other hand, $\Gamma_{n}\in STB_{n}\Leftrightarrow\mathcal{G}(\Gamma_{n})=4$, i.e. it has no 3-cycle. Thus no two adjacent vertices share an adjacent vertex.

        $\forall\Gamma_{n}\in STB_{n}$, let's denote $u_{1}^{i}$ and $u_{1}^{j}$ $(i\ne j)$ for any two non-adjacent vertices. Now we choose any other pair index $k$ which is different from $i$ and $j$. $u_{1}^{i}$ need to be adjacent to either $u_{1}^{k}$ or $u_{2}^{k}$, and so as $u_{1}^{j}$. $u_{1}^{i}$ and $u_{1}^{j}$ can't be adjacent to $u_{1}^{k}$ and $u_{2}^{k}$ respectively since there will be a logical error in the graph $\Omega(\Gamma_{n},u_{1}^{i}u_{2}^{i}u_{1}^{j}u_{2}^{j})$, which leads to $\Gamma_{n}\notin STB_{n}$ as shown in Figure \ref{Figure 14}.

        Thus $\forall\Gamma_{n}\in STB_{n}$, any two non-adjacent vertices have $n$ common adjacent vertices derived from $n$ pairs of vertices (or labels).

        Consequently, for any $\Gamma_{n}\in STB_{n}$, we have $\Gamma_{n}$ is isomorphic to $srg(2n,n,0,n)$. Furthermore, we know they are equivalent if we consider the circumstance of $\Gamma_{n}\notin STB_{n}$ again.

        \qedhere
      \end{proof}

      \begin{description}
	    \item[Lemma 28 \citep{Bollobas1998P274}]  For any strongly regular graph
          $\Gamma=srg(n,k,p,q)$, we have
        \begin{equation*}
          A(\Gamma)^2+(q-p)A(\Gamma)=(k-q)I_{n}+qJ_{n}.
        \end{equation*}
      \end{description}

      \begin{proof}
        Suppose $V(\Gamma)=\{v_{1},v_{2},\dots,v_{n}\}$. We have the matrix $A(\Gamma)$, denoted as $\{a_{i,j}\}_{n\times n}$, and its square $A(\Gamma)^2$, denoted $\{b_{i,j}\}_{n\times n}$.
        Then, through the multiplication of two matrices, we have
        \begin{equation*}
          \label{Formula6}
          b_{i,j}=\sum_{t=1}^n{a_{i,t}a_{t,j}}  \quad\quad\quad \forall 1\ge i,j\ge n.\tag{6}
        \end{equation*}
        
        \begin{figure}
        	\centering
        	\includegraphics[height=4cm]{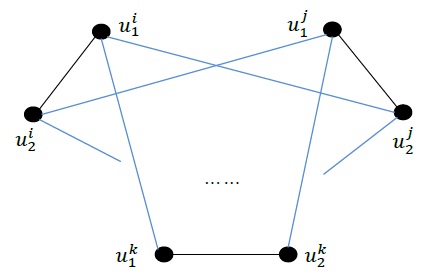}
        	\caption{The graph $\Gamma_{n}$ with $u_{1}^{i}$ and $u_{1}^{j}$ being adjacent to $u_{1}^{k}$ and $u_{2}^{k}$ respectively. Thus in the graph $\Omega(\Gamma_{n},u_{1}^{i}u_{2}^{i}u_{1}^{j}u_{2}^{j}),u_{1}^{k}$ and $u_{2}^{k}$ share a common vertex $\{u_{1}^{i},u_{1}^{j}\}$ (or $u_{1}^{min\{i,j\}}$ equivalently).}
        	\label{Figure 14} 
        \end{figure}

        Noticing that for every vertex $v_{t}$ in the Formula (\ref{Formula6}),
        \begin{equation*}
          a_{i,t}a_{t,j}=
          \left\{
             \begin{array}{lr}
             1 & (v_{t}\mbox{ is adjacent to both }v_{i}\mbox{ and }v_{j}) \\
             0 & \mbox{(Otherwise)}
             \end{array}
          \right.
        \end{equation*}

        hence $b_{i,j}$ is the number of 2-path from $v_{i}$ to $v_{j}$. Here 2-path is a path with a length of 2.

        If $v_{i}v_{j}\in E(\Gamma)$, i.e. $a_{i,j}=a_{j,i}=1$, the number of 2-path from $v_{i}$ to $v_{j}$ is the same as the number of common adjacent vertices of $v_{i}$ and $v_{j}$, which is $p$. Thus $b_{i,j}=p$ and we have
        \begin{equation*}
          \label{Formula7}
          b_{i,j}+(q-p)a_{i,j}=q \quad\quad\quad \forall v_{i}v_{j}\in E(\Gamma).\tag{7}
        \end{equation*}

        Apparently, all $v_{i}$ and $v_{j}$ in the Formula (\ref{Formula7}) satisfy $i\ne j$, which means it won't occur on the diagonal of matrices.

        If $v_{i}v_{j}\notin E(\Gamma)$ and $i\ne j$, i.e. $a_{i,j}=a_{j,i}=0 \quad (i\ne j)$, the number of 2-path from $v_{i}$ to $v_{j}$ is the same as $q$ (i.e. the number of common adjacent vertices of $v_{i}$ and $v_{j}$). Thus $b_{i,j}=q$ and we have
        \begin{equation*}
          \label{Formula8}
          b_{i,j}+(q-p)a_{i,j}=q \quad\quad\quad \forall v_{i}v_{j}\notin E(\Gamma)\mbox{ and }i\ne j.\tag{8}
        \end{equation*}

        If $i=j$, the number of 2-path from $v_{i}$ to $v_{i}$ is the same as the degree of $v_{i}$, which is $k$. Thus $b_{i,i}=k$ and we have, since $a_{i,i}=0$,
        \begin{equation*}
          \label{Formula9}
          b_{i,i}+(q-p)a_{i,i}=k \quad\quad\quad \forall i\in [1,n].\tag{9}
        \end{equation*}

        From Formulas (\ref{Formula7}) to (\ref{Formula9}), we finally have
        \begin{equation*}
          A(\Gamma)^2+(q-p)A(\Gamma)=(k-q)I_{n}+qJ_{n}.
        \end{equation*}

        \qedhere
      \end{proof}

      \begin{proof}[Proof of Theorem 25 (Matrix Criterion)]
         \quad

         For any bipartitely stable stereotype graph $\Gamma_{n}\in STB_{n}$, we know $\Gamma_{n}$ is isomorphic to $srg(2n,n,0,n)$ by Lemma 22. Then by Lemma 23, we have $A(\Gamma_{n})^2+nA(\Gamma_{n})=nJ_{2n}.$

         On the other hand, consider $\Gamma_{n}\notin STB_{n}$. According to the proof of Lemma 22, there exist two non-adjacent vertices from different pairs of labels, denoted $u_{1}^{i}$ and $u_{1}^{j}$ where $i\ne j$, who are adjacent to $u_{1}^{k}$ and $u_{2}^{k}$ respectively where $k$ is a pair index different to $i$ and $j$. Then the number of common adjacent vertices of $u_{1}^{i}$ and $u_{1}^{j}$ is less than $n$, i.e. the number of 2-path from $u_{1}^{i}$ to $u_{1}^{j}$ is less than $n$.

         Suppose $u_{1}^{i}$ is the $r$th vertex of $\Gamma_{n}$, while $u_{1}^{j}$ is the $s$th vertex of $\Gamma_{n}$. Borrowing notations from the proof of Lemma 23, we have
         \begin{equation*}
           \label{Formula10}
           b_{r,s}+na_{r,s}<n+n\times 0=n.\tag{10}
         \end{equation*}

         The left side of Formula (\ref{Formula10}) is the element in the $r$th row and the $s$th column of the matrix $A(\Gamma_{n})^2+nA(\Gamma_{n})$. Since this element is less than $n$, we have $A(\Gamma_{n})^2+nA(\Gamma_n)\ne nJ_{2n}$.

         To sum up, $\forall \Gamma_{n}\in ST_{n} (n\ge2)$, we have
         \begin{equation*}
           \Gamma_{n}\in STB_{n}\Leftrightarrow A(\Gamma_{n})^2+nA(\Gamma_{n})=nJ_{2n}.
         \end{equation*}

         which is the matrix criterion.

         \qedhere
      \end{proof}

    \subsection{The Characteristic Polynomial and Stereotype Graphs}
      We now explore the characteristic polynomial of the adjacency matrix of a stereotype graph.
      \begin{description}
	    \item[Definition 29]  The \emph{characteristic polynomial} of a graph $\Gamma$ with $n$ vertices is defined as $f(\Gamma,\lambda)=det(\lambda I_{n}-A(\Gamma))=\sum_{i=0}^n c_{i}{\lambda}^{n-i}$, where $I_{n}$ is a $n\times n$ identity matrix.
      \end{description}

      \begin{description}
	    \item[Proposition 30 \citep{Zhan2013}]  For every positive integers $i$, $(-1)^i c_{i}$ is the sum of all $i\times i$ principal minors of $A(\Gamma)$.
      \end{description}

      Since every $\Gamma_{n}\in ST_{n}$ has $2n$ vertices, its characteristic polynomial is
      \begin{equation*}
        \label{Formula11}
        f(\Gamma_{n},\lambda)=det(\lambda I_{2n}-A(\Gamma_{n}))=\sum_{i=0}^{2n} c_{i}{\lambda}^{2n-i}.\tag{11}
      \end{equation*}

      As we can efficiently obtain the characteristic polynomial of a matrix computationally, it would be better if the bipartite stability can be derived from the related characteristic polynomial. Fortunately, its coefficients $c_{i}$ have certain regularity.
      \begin{description}
	    \item[Theorem 31]  $\forall \Gamma_{n}\in ST_{n}$ $(n\ge 2)$, in Formula (\ref{Formula11}) we have
        \begin{enumerate}
          \item $c_{0}=1$;
          \item $c_{1}=0$;
          \item $c_{2}=-n^2$.
        \end{enumerate}
      \end{description}

      \begin{proof}
        (1) The first statement is apparent according to Definition 29.

        (2) By Proposition 30, $c_{1}=tr(A(\Gamma_{n}))$ where $tr(M)$ is the trace of the matrix $M$. Since all diagonal elements of an adjacency matrix are $0$, we have $c_{1}=0$.

        (3) All possible types of $2\times 2$ principal minors of $A(\Gamma_{n})$ are $det(\begin{bmatrix}0&0\\0&0\end{bmatrix})=0$ and $det(\begin{bmatrix}0&1\\1&0\end{bmatrix})=-1$. The former one indicates two non-adjacent vertices while the latter one indicates two adjacent vertices, i.e. an edge.

        Thus by Proposition 30 and Proposition 6(1), we have
        \begin{equation*}
            c_{2}=-2E(\Gamma_{n})=-n^2.
        \end{equation*}

        \qedhere
      \end{proof}

      The following criterion shows how bipartite stability emerges directly from the coefficient of the characteristic polynomial. It allows us to judge the bipartite stability computationally, just like the matrix criterion.
      \begin{description}
	    \item[Theorem 32 (Characteristic Criterion)]  \quad

          $\forall\Gamma_{n}\in ST_{n} (n\ge 2),$
          \begin{equation*}
            \Gamma_{n}\in STB_{n}\Leftrightarrow c_{3}=0,
          \end{equation*}
          where $c_{3}$ is the related coefficient in Formula (\ref{Formula11}).
      \end{description}

      \begin{proof}
        All possible types of $3\times 3$ principal minors of $A(\Gamma_{n})$ are
        \begin{equation*}
          det(\begin{bmatrix}0&1&1\\1&0&0\\1&0&0\end{bmatrix})=0,
          det(\begin{bmatrix}0&1&0\\1&0&0\\0&0&0\end{bmatrix})=0,
        \end{equation*}
        \begin{equation*}
          det(\begin{bmatrix}0&0&1\\0&0&0\\1&0&0\end{bmatrix})=0,
          det(\begin{bmatrix}0&0&0\\0&0&1\\0&1&0\end{bmatrix})=0,
        \end{equation*}
        \begin{equation*}
          det(\begin{bmatrix}0&1&0\\1&0&1\\0&1&0\end{bmatrix})=0,
          det(\begin{bmatrix}0&0&1\\0&0&1\\1&1&0\end{bmatrix})=0,
        \end{equation*}
        and
        \begin{equation*}
          det(\begin{bmatrix}0&1&1\\1&0&1\\1&1&0\end{bmatrix})=2,
        \end{equation*}

        where only the last one is equivalent to a 3-cycle.

        Thus by Proposition 30, $-c_{3}/2$ is the number of 3-cycles contained in $\Gamma_{n}$. Then by the girth criterion, we have, $\forall \Gamma_{n}\in ST_{n} (n\ge 2)$,
         \begin{equation*}
           \Gamma_{n}\in STB_{n}\Leftrightarrow c_{3}=0.
         \end{equation*}

        \qedhere
      \end{proof}

      \begin{description}
	    \item[Corollary 33]  $\forall\Gamma_{n}\in ST_{n} (n\ge 2), c_{3}\le 0$ and $c_{3}/4$ is an integer.
      \end{description}
  
  	  \begin{figure}
  		\centering
  		\includegraphics[height=10cm]{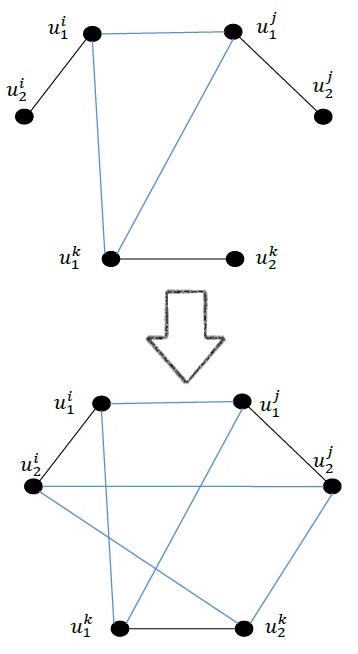}
  		\caption{Every three pairs of vertices(labels) can not contain just one 3-cycle.}
  		\label{Figure 15}
  	  \end{figure}
      
      \begin{proof}
        According to the proof of Theorem 31, $-c_{3}/2$ is the number of 3-cycles contained in $\Gamma_{n}$, denoted $\eta$. Since $\eta$ is not negative, we have $c_{3}=-2\eta\le0$.

        Suppose a 3-cycle, denoted $u_{1}^{i}u_{1}^{j}u_{1}^{k}$, is contained in $\Gamma_{n}$. Then another 3-cycle $u_{2}^{i}u_{2}^{j}u_{2}^{k}$ is also contained in $\Gamma_{n}$ as shown in Figure \ref{Figure 15}.

        Thus, every three pairs of vertices(labels) contains zero or two 3-cycles, i.e. the number of three pairs of vertices(labels) who contains 3-cycles is $-c_{3}/4$. Thus $c_{3}/4$ is an integer.

        \qedhere
      \end{proof}
  
    \subsection{The Chromatic Polynomial and Stereotype Graphs}
      Recall the definition of $k$-colorable graphs (Definitions 13 and 14) in the vertex coloring again. The number of vertex colorings with finite colors of a graph is always an interesting and challenging question in mathematics. It has been proved that the number of vertex colorings of a graph can be represented by a polynomial \citep{Birkhoff1912}.

      Given a graph $\Gamma$ with $n$ vertices and a number of different colors $x$, we can easily know its number of vertex $x$-colorings, denoted $C(\Gamma,x)$.
      \begin{description}
	    \item[Theorem 34 \citep{Biggs1974P63}] We have
          \begin{equation*}
            \label{Formula12}
            C(\Gamma,x)=\sum_{i=0}^n (m_{i}(\Gamma)\prod_{j=0}^{i-1}(x-j))\tag{12}
          \end{equation*}

          where $m_{i}(\Gamma)$ is the number of vertex coloring $i$-partitions of $\Gamma$.
      \end{description}

      \begin{proof}
        For every $i\in[0,n], m_{i}(\Gamma)\prod_{j=0}^{i-1}(x-j)$ is the number of vertex $x$-colorings of $\Gamma$ which just uses $i$ colors. Thus $\sum_{i=0}^n(m_{i}(\Gamma)\prod_{j=0}^{i-1}(x-j))$ is the number of vertex $x$-colorings of $\Gamma$, which is $C(\Gamma,x)$.

        \qedhere
      \end{proof}

      \begin{description}
	    \item[Corollary 35]  $\Gamma$ is $k$-colorable$ \Leftrightarrow C(\Gamma,k)>0$; $\Gamma$ is not $k$-colorable$ \Leftrightarrow C(\Gamma,k)=0$.

        \item[Corollary 36]  $C(\Gamma,x)$ is a polynomial of $x$ with degree $n$.
      \end{description}

      \begin{proof}
        This is because $m_{n}(\Gamma)=1$.

        \qedhere
      \end{proof}

      \begin{description}
        \item[Definition 37 \citep{Dong2005}]
        The \emph{chromatic number} of a graph $\Gamma$ is defined as
        \begin{equation*}
          \chi(\Gamma)=min\{x|x\in\mathbb{Z}_{+},C(\Gamma,x)\ge 1\}.
        \end{equation*}
      \end{description}

      In fact, the chromatic number of a graph $\Gamma$ is the smallest number of colors which can be used to color $\Gamma$.

      Now we start to try to use chromatic numbers and chromatic polynomials to judge the bipartite stability.

      \begin{description}
        \item[Theorem 38]  $\forall\Gamma_{n}\in ST_{n}$, the following three statements are equivalent:
        \begin{enumerate}
          \item $\Gamma_{n}\in STB_{n}$;
          \item $\chi(\Gamma_{n})=2$;
          \item $C(\Gamma,2)>0$.
        \end{enumerate}
      \end{description}

      \begin{proof}
        This is apparent from Theorem 15 and Corollary 35.

        \qedhere
      \end{proof}

      \begin{description}
        \item[Theorem 39]  $\forall\Gamma_{n}\in ST_{n}$, The chromatic polynomial of $\Gamma_{n}$ can be represented as
        \begin{equation*}
          \label{Formula13}
          C(\Gamma_{n},x)=\sum_{i=0}^{2n} b_{i}x^{2n-i}.\tag{13}
        \end{equation*}
      \end{description}

      \begin{proof}
        By Corollary 36, we know that the degree of $C(\Gamma_n,x)$ is $2n$.

        \qedhere
      \end{proof}

      In common with the characteristic polynomial, we can computationally obtain the chromatic polynomial of a graph from its adjacency matrix. Thus it would be better if the bipartite stability is related to the chromatic polynomial directly. Fortunately, we have the following criterion.

      \begin{description}
        \item[Theorem 40 (Chromatically Bipartite Criterion)] \quad

        $\forall\Gamma_{n}\in ST_{n}$ $(n\ge 2)$, we have
        \begin{equation*}
          \Gamma_{n}\in STB_{n}\Leftrightarrow b_{2}=\binom{n^2}{2}
        \end{equation*}

        as well as
        \begin{equation*}
          \Gamma_{n}\notin STB_{n}\Leftrightarrow b_{2}<\binom{n^2}{2}.
        \end{equation*}
      \end{description}

       We introduce a necessary lemma before we give the proof of Theorem 40. The lemma succinctly describes the coefficients of chromatic polynomials of certain graphs. The complete proof of which is complicated and can be found in \citet{Biggs1974P7376}, conducted with the help of rank polynomials.

      \begin{description}
        \item[Lemma 41 \citep{Biggs1974P7376}]  Suppose in a graph $\Gamma$, the number of contained $\mathcal{G}(\Gamma)$-cycles is $\eta$. Using notations above, we have
        \begin{enumerate}
          \item $(-1)^{i}b_{i}=\binom{|E(\Gamma)|}{i}$ where $i=0,1,\dots,\mathcal{G}(\Gamma)-2$ and
          \item $(-1)^{i}b_{i}=\binom{|E(\Gamma)|}{i}-\eta$ where $i=\mathcal{G}(\Gamma)-1$.
        \end{enumerate}
      \end{description}

      \begin{proof}[Proof of Theorem 40]
        \emph{(Chromatically Bipartite Criterion)}

        By the girth criterion, we have $\Gamma_{n}\in STB_{n}\Leftrightarrow \mathcal{G}(\Gamma)=4$ and $\Gamma_{n}\notin STB_{n}\Leftrightarrow \mathcal{G}(\Gamma)=3$.

        By $\Gamma_{n}\in STB_{n}\Leftrightarrow \mathcal{G}(\Gamma)=4$ and Lemma 41, we have, $\forall\Gamma_{n}\in STB_{n}$,
        \begin{equation*}
          (-1)^i b_{i}=\binom{n^2}{i} \quad (i=0,1,2).
        \end{equation*}

        i.e. $\forall\Gamma_{n}\in STB_{n}$,
        \begin{equation*}
          \label{Formula14}
          b_{0}=1, b_{1}=-n^2, b_{2}=\binom{n^2}{2}.\tag{14}
        \end{equation*}

        By $\Gamma_{n}\notin STB_{n}\Leftrightarrow\mathcal{G}(\Gamma)=3$ and Lemma 41, we have, $\forall\Gamma_{n}\notin STB_{n}$,
        \begin{equation*}
          (-1)^i b_{i}=\binom{n^2}{i} \quad (i=0,1).
        \end{equation*}
        \quad\quad\quad and

        \begin{equation*}
          (-1)^i b_{i}=\binom{n^2}{i}-\eta \quad (i=2).
        \end{equation*}

        where $\eta$ is the number of 3-cycles contained in $\Gamma_{n}$.

        Since $\Gamma_{n}\notin STB_{n}$, $\eta$ is positive. Thus, $\forall\Gamma_{n}\notin STB_{n}$,
        \begin{equation*}
          \label{Formula15}
          b_{0}=1, b_{1}=-n^2, b_{2}<\binom{n^2}{2}.\tag{15}
        \end{equation*}

        By Formulas (\ref{Formula14}) and (\ref{Formula15}), we prove the criterion.

        \qedhere
      \end{proof}

      \begin{description}
        \item[Corollary 42]  $\forall\Gamma_{n}\in ST_{n}$,
        \begin{equation*}
          b_{0}=1, b_{1}=-n^2, b_{2}\le\binom{n^2}{2}.
        \end{equation*}
      \end{description}

      \begin{proof}
        This is due to Formulas (\ref{Formula14}) and (\ref{Formula15}).

        \qedhere
      \end{proof}

      \begin{description}
        \item[Corollary 43]  $\forall\Gamma_{n}\in ST_{n}$ $(n\ge2), b_{2}=\binom{n^2}{2}+c_{3}/2$, where $c_{3}$ is the coefficient of the characteristic polynomial of $A(\Gamma_{n})$ in Formula (\ref{Formula11}).
      \end{description}

      \begin{proof}
        According to the proof of Theorem 31, $-c_{3}/2$ is the number of 3-cycles contained in $\Gamma_{n}$, denoted $\chi$.

        Then by Lemma 41, we have $b_{2}=\binom{n^2}{2}-\chi=\binom{n^2}{2}+c_{3}/2$.

        \qedhere
      \end{proof}

  \section{Summary of Criteria for Bipartite Stability}
    In this section, we provide a list of criteria for judging the bipartite stability to summarize the discussion on stereotype graphs based on complementary labels.

    \begin{description}
	    \item[Coloring Criterion (Theorem 15)]  \quad

        $\forall\Gamma_{n}\in ST_{n}$, we have
        \begin{center}
          $\Gamma_{n}\in STB_{n}\Leftrightarrow \Gamma_{n}$ is 2-colorable.
        \end{center}

        \item[Bipartite Criterion (Theorem 17)]  \quad

          $\forall\Gamma_{n}\in ST_{n}$, we have
          \begin{center}
            $\Gamma_{n}\in STB_{n}\Leftrightarrow \Gamma_{n}$ is isomorphic to $K_{n,n}$.
          \end{center}

        \item[Girth Criterion (Theorem 20)]  \quad

          $\forall\Gamma_{n}\in ST_{n}$ $(n\ge2)$, we have
          \begin{equation*}
            \Gamma_{n}\in STB_{n}\Leftrightarrow\mathcal{G}(\Gamma_{n})=4.
          \end{equation*}

        \item[Matrix Criterion (Theorem 25)]  \quad

           $\forall\Gamma_{n}\in ST_{n}$ $(n\ge2)$, we have
           \begin{equation*}
             \Gamma_{n}\in STB_{n}\Leftrightarrow A(\Gamma_{n})^2+nA(\Gamma_{n})=nJ_{2n}.
           \end{equation*}

        \item[Characteristic Criterion (Theorem 32)]  \quad

          $\forall\Gamma_{n}\in ST_{n}$ $(n\ge2),$
          \begin{equation*}
            \Gamma_{n}\in STB_{n}\Leftrightarrow c_{3}=0,
          \end{equation*}

        \item[Chromatically Bipartite Criterion (Theorem 40)]  \quad

        $\forall\Gamma_{n}\in ST_{n}$ $(n\ge2)$, we have
        \begin{equation*}
          \Gamma_{n}\in STB_{n}\Leftrightarrow b_{2}=\binom{n^2}{2},
        \end{equation*}
     \end{description}

     \quad

     In comparison to the former three criteria, the latter three criteria are more convenient to use computationally.

     It is worth mentioning the coloring criterion again before we turn to discuss stereotype graphs based on contradictory labels. On the one hand, the coloring criterion has a unique psychological meaning, as we mentioned after the statement of coloring criterion. On the other hand, the psychological meaning of ``color'' may change according to the psychological meaning that stereotype graphs hold originally, as we will notice in the next section.

  \section{Stereotype Graphs based on Contradictory Labels}

      We now give a different interpretation of stereotype graphs. Recall the example of a stereotype based on two pairs of complementary labels. In Figure \ref{Figure 4}, we can regard ``Femininity'' and ``Beneficence'' as identical labels. However, such a conclusion is based on the assumption that every two labels in the same pair are complementary. Once Anne only thinks that ``Beneficence'' and ``Severity'' (two labels in the same pair) are contradictory, Formula (\ref{Formula1}) (i.e. $\mathcal{F}\subseteq\mathcal{B}$) would become uncertain. Consequently, she can't regard ``Femininity'' and ``Beneficence'' as identical labels, which makes the merge operations unreasonable to act on the stereotype graph.

      Despite the invalidation of merge operations, Figure \ref{Figure 4} can persist in Anne's cognition rationally. Suppose $v_{1}$ and $v_{2}$, as well as $u_{1}$ and $u_{2}$, are no longer complementary but just contradictory. If we still use vertices to represent labels, every two adjacent vertices represent the corresponding contradictory labels, i.e. everyone can be categorized into neither of these two labels.

      Another problem is the reasonableness of stereotype graphs like Figure \ref{Figure 3}. Although ``Femininity'' and ``Masculinity'' are not complementary anymore, they are still in the same pair of labels, which indicates a property of one's personality. If ``Severity'' is both contradictory with ``Femininity'' and ``Masculinity'', Anne may wonder whether these two pairs of labels are correlated with each other. Then a reasonable consequence is that Anne admits her cognitive thought here is suspicious and thus compromises to amend her stereotype.

      Thus, we can make a reasonable assumption that in stereotype graphs based on contradictory labels, any two pairs of vertices (labels) as an induced subgraph don't contain any 3-cycles, i.e. it is a basic stereotype induced subgraph. This is the same as the circumstance of stereotype graphs based on complementary labels. Therefore, when we add pairs of contradictory labels to form a stereotype graph, we can still use the definition of stereotype graphs based on complementary graphs (see Definition 3 and Notation 4).

      Notice that although stereotype graphs are the same as before mathematically, they have totally different meanings. Every two adjacent vertices indicate that they are contradictory labels now. Moreover, the merge operation is invalid henceforth, which asks for a new way to judge the stability of a stereotype graph.

    \subsection{Chromatic Stability and CSI Index}
      The new way to estimate the stability of a stereotype graph is enlightened by \emph{stereotyping process} and \emph{vertex coloring}. The idea is based on \emph{stereotype preservation biases}, which refers to the tendency that we prefer confirming information to disconfirming information when we test our stereotypical beliefs \citep{Johnson1996}.

      Suppose Anne has already met several people. Since she is about to judge their personalities or characteristics, according to her stereotypes unconsciously, in order to decide whom she'd better make friends with, she starts to categorize people by putting related ``labels'' on them. This is actually a stereotyping process, the third component of a stereotype mentioned by \citet{Perry2000a}.

      If we view different \emph{people} as different colors, then the stereotyping process above can be viewed as an attempt of vertex coloring, since two adjacent vertices (labels) are contradictory and should belong to different people, i.e. different colors.

      Of course, we have to raise some reasonable hypotheses to ensure that the procedure is a vertex coloring. Every label should belong to at least to one person, or the label which belongs to no people has no meanings in the stereotype. In addition, every label belongs to at most one person to reduce the possibility of the failure of the procedure.

      With the rational hypothesis above, the stereotyping process now is actually an attempt at vertex coloring, where the vertices represent labels and colors represent people. We will call it the \emph{coloring process} afterward.

      For a certain stereotype graph, a coloring process can either success or fail. With the idea of stereotype preservation biases, we expect the coloring process to be as far as possible so that the present stereotype can still be preserved. However, if there are too few people, colors will be not enough to color the stereotype graph, which induces the inevitable failure of the coloring procedure. Once the ineluctability of such failure (we call it the \emph{inevitable error}) is detected by Anne herself, she will then realize that she holds an irrational stereotype, which may impel Anne to modify it or even abandon it.

      In general, the \emph{inevitable error} in the stereotype is the inevitable failure of conducting coloring process. It is also fatal, since it also indicates the present stereotype is nonsense, as the logical error does. Once such kind of error is detected in the stereotype, the modification becomes essential. In another word, a stereotype with inevitable errors is hard to linger on cognition without change, which makes it unstable.

      Hence, as the analysis above, the \emph{stability} of a stereotype graph is corresponded to the existence of inevitable errors, which is directly related to the number of people (or colors) and the structure of the graph. More people, i.e. more usable colors, can make the stereotype graph more easily stay in one's cognition without inevitable errors.

      Therefore, for a certain stereotype graph, we judge its stability by detecting its smallest number of necessary colors which can make the coloring process success possibly. By Definition 37, we know that it is actually the chromatic number of the stereotype graph.

      \begin{description}
	    \item[Definition 44] $\forall\Gamma_{n}\in ST_{n}$, the chromatic number of $\Gamma_{n}$, denoted $\chi(\Gamma_{n})$, is the \emph{chromatic stability index} (\emph{CSI}) of $\Gamma_{n}$. Such property is called \emph{chromatic stability}.

        \item[Definition 45 (Chromatic Criterion)]     \quad

        $\forall\Gamma_{n}^{(1)}\in ST_{n}$ and $\Gamma_{m}^{(2)}\in ST_{m}$:
        \begin{enumerate}
          \item $\Gamma_{n}^{(1)}$ is more \emph{chromatically stable} (or \emph{stable}) to $\Gamma_{m}^{(2)}$ if
            \begin{equation*}
              \chi(\Gamma_{n}^{(1)})<\chi(\Gamma_{m}^{(2)});
            \end{equation*}
          \item $\Gamma_{n}^{(1)}$ is more \emph{chromatically unstable} (or \emph{unstable}) to $\Gamma_{m}^{(2)}$ if
            \begin{equation*}
              \chi(\Gamma_{n}^{(1)})>\chi(\Gamma_{m}^{(2)});
            \end{equation*}
          \item $\Gamma_{n}^{(1)}$ is the \emph{same chromatically stable} (or \emph{same stable}) as $\Gamma_{m}^{(2)}$ if
            \begin{equation*}
              \chi(\Gamma_{n}^{(1)})=\chi(\Gamma_{m}^{(2)});
            \end{equation*}
        \end{enumerate}

        \item[Notation 46] Set $ST_{n}(k)=\{\Gamma_{n}|\Gamma_{n}\in ST_{n}$ and $\chi(\Gamma_{n})=k\}$.
      \end{description}

      Definition 45 is a new way to classify different stereotype graphs through CSI. It is deeply correlated with bipartite stability.
      \begin{description}
	    \item[Theorem 47] $ST_{n}(2)=STB_{n}$.
      \end{description}

      \begin{proof}
        It is straightforward from the color criterion and Notation 46.

        \qedhere
      \end{proof}

      From Theorem 47, we conclude that chromatic stability is a more generalized judgment compared to bipartite stability.

    \subsection{The Precise Range of CSI}
      After we define chromatic stability, the extent of the stability of a stereotype graph can be considered as a number, CSI, in a certain range. Here we attempt to detect the range of CSI.
      \begin{description}
	    \item[Theorem 48] $\forall\Gamma_{n}\in ST_{n}$ $(n\ge2),2\le\chi(\Gamma_{n})\le n$.
      \end{description}

\begin{figure}
      	  \centering
          \includegraphics[height=4cm]{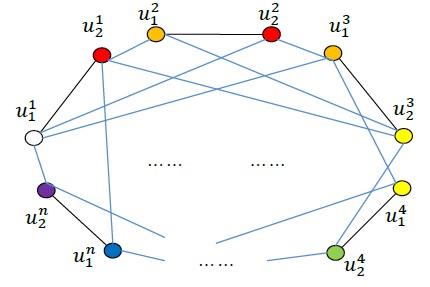}
          \caption{An example of the vertex coloring after all operations with the color of $u_{1}^{1}$ remaining.}
          \label{Figure 16}
        \end{figure}

      \begin{proof}
        Since $E(\Gamma_{n})\ne\emptyset$, we have $\chi(\Gamma_{n})\ge2$.

        If $n=2$, we have $ST_{2}=\{K_{2,2}\}$ by Corollary 18. Thus $\chi(\Gamma_{2})=2$.

        If $n\ge3$, $\Gamma_{n}$ is $n$-regular by Proposition 6(5). Next we give a coloring procedure to ensure that $\Gamma_{n}$ is $n$-colorable, i.e. $\chi(\Gamma_{n})\le n$.

        $\Gamma_{n}$ has $n$ pairs of vertices denoted by $u_{1}^{i}$ and $u_{2}^{i}$, where $i\in[1,n]$ and $i\in\mathbb{Z}_{+}$. A vertex $n$-coloring is denoted by a mapping $\theta:V(\Gamma_{n})\rightarrow\{1,2,\dots,n\}$ where $\theta(u_{p}^{i})\ne\theta(u_{q}^{j})$ for any $u_{p}^{i}u_{q}^{j}\in E(\Gamma_{n})$.

        Let $\theta(u_{2}^{1})=1$,

        \begin{equation*}
          \theta(u_{1}^{2})=
          \left\{
             \begin{array}{lr}
               1 \quad (u_{2}^{1}u_{1}^{2}\notin E(\Gamma_{n})) &  \\
               2 \quad (u_{2}^{1}u_{1}^{2}\in E(\Gamma_{n})) &
             \end{array}
          \right.
        \end{equation*}

        and $\theta(u_{2}^{2})=3-\theta(u_{1}^{2})$. We can easily see that the coloring among $u_{2}^{1},u_{1}^{2}$ and $u_{2}^{2}$ is reasonable.

        For $i\in\{3,4,\dots,n\}$, we do the similar operation in sequence, i.e.

        \begin{equation*}
          \theta(u_{1}^{i})=
          \left\{
             \begin{array}{lr}
               i-1 \quad & (u_{1}^{i}u_{p}^{i-1}\notin E(\Gamma_{n}))  \\
               i \quad & (u_{1}^{i}u_{p}^{i-1}\in E(\Gamma_{n}))
             \end{array}
          \right.
        \end{equation*}

        where $p$ satisfies $\theta(u_{p}^{i-1})=i-1$, and
        \begin{equation*}
          \theta(u_{2}^{i})=(2i-1)-\theta(u_{1}^{2}).
        \end{equation*}

         As for each operation, the color $i$ has never occurred before, as well as the color $i-1$ just occurs on two non-adjacent vertices respectively located in the $i$th pair and the $(i-1)$th pair of vertices. In this way, the coloring among the former $i$ pairs of vertices (except for $u_{1}^{1}$) is indeed a vertex coloring. An example is shown in Figure \ref{Figure 16}.

        The only problem is the color of $u_{2}^{1}$. If $u_{1}^{1}u_{1}^{2}\in E(\Gamma_{n})$, then $u_{2}^{1}u_{1}^{2}\notin E(\Gamma_{n})$. Thus $\theta(u_{1}^{2})=1=\theta(u_{2}^{1})$. On the other hand, if $u_{1}^{1}u_{2}^{2}\in E(\Gamma_{n})$, then $u_{2}^{1}u_{1}^{2}\in E(\Gamma_{n})$. Thus $\theta(u_{1}^{2})=2$ and $\theta(u_{2}^{2})=1=\theta(u_{2}^{1})$.

        Hence, $u_{1}^{1}$ is always adjacent to two vertices with the same color. Since $\Gamma_{n}$ is $n$-regular, the degree of $u_{1}^{1}$ is $n$. Considering there are just $n$ colors, we know the fact that there exists a color, denoted $k$, belonging to no vertices that are adjacent to $u_{1}^{1}$. Let $\theta(u_{1}^{1})=k$, and we have a vertex coloring finally.

        Thus $\Gamma_{n}$ is $n$-colorable, i.e. $\chi(\Gamma_{n})\le n$.

        \qedhere
      \end{proof}

      Theorem 48 shows a range of CSI that only applies to $n\ge2$. When $n=1$, we have $\chi(\Gamma_{1})=2$. The following two theorems indicate that 2 and $n$ are always the infimum and supremum of CSI of $\Gamma_{n}$ respectively, where $n\ge2$.
      \begin{description}
	    \item[Theorem 49] $\forall n\in \mathbb{Z}_{+}, ST_{n}(2)\ne\emptyset$.
      \end{description}

      \begin{proof}
	    By Theorem 44, we have $ST_{n}(2)=STB_{n}$.

        Let $u_{1}^{i}$ and $u_{2}^{i}$ $(i\in\{1,2,\dots,n\})$ denote $n$ pairs of vertices. These vertices form a stereotype graph $\Gamma_{n}$.

        Denote $\mathcal{U}=\{u_{1}^{i}|i\in[1,n],i\in\mathbb{Z}_{+}\}$ and $\mathcal{V}=\{u_{2}^{i}|i\in[1,n],i\in\mathbb{Z}_{+}\}$. Then $\mathcal{U}$ and $\mathcal{V}$ are a partition of $V(\Gamma_{n})$.

        Now define $E(\Gamma_{n})=\{ab|a\in\mathcal{U},b\in\mathcal{V}\}$. It is obvious that $\Gamma_{n}\in ST_{n}$ and $\Gamma_{n}$ is isomorphic to $K_{n,n}$. By the bipartite criterion, we have $\Gamma_{n}\in STB_{n}=ST_{n}(2)$.

        \qedhere
      \end{proof}

      \begin{description}
	    \item[Theorem 50] $\forall n\in \mathbb{Z}_{+}$ and $n\ge2, ST_{n}(n)\ne\emptyset$.
      \end{description}

      \begin{proof}
	    Let $u_{1}^{i}$ and $u_{2}^{i}$ $(i\in\{1,2,\dots,n\})$ denote $n$ pairs of vertices. These vertices form a stereotype graph $\Gamma_{n}$.

        Denote $\mathcal{U}=\{u_{1}^{i}|i\in[1,n],i\in\mathbb{Z}_{+}\}$ and $\mathcal{V}=\{u_{2}^{i}|i\in[1,n],i\in\mathbb{Z}_{+}\}$. Then $\mathcal{U}$ and $\mathcal{V}$ are a partition of $V(\Gamma_{n})$.

        Now let
        \begin{eqnarray*}
          \mathfrak{S}&=&\{u_{1}^{i}u_{2}^{i}|i\in[1,n]\},\\
          \mathfrak{U}&=&\{ab|a\in\mathcal{U},b\in\mathcal{U},a\ne b\}, \\
          \mathfrak{V}&=&\{ab|a\in\mathcal{V},b\in\mathcal{V},a\ne b\},\mbox{ and} \\
          E(\Gamma_{n})&=&\mathfrak{S}\cup\mathfrak{U}\cup\mathfrak{V}.
        \end{eqnarray*}

        It is straightforward to show that $\Gamma_{n}\in ST_{n}$. Since $|\mathfrak{U}|=|\mathfrak{V}|=n$, $\mathfrak{U}$ and $\mathfrak{V}$ indicate two $K_{n}$s, respectively. According to $\mathfrak{U}\subseteq E(\Gamma_{n})$, we have $\chi(\Gamma_{n})\ge n$.

        Thus, by Theorem 48, we have $\chi(\Gamma_{n})=n$ and $\Gamma_{n}\in ST_{n}(n)$.

        \qedhere
      \end{proof}

      $\Gamma_{n}$ constructed in the proof of Theorem 50 is just like two complete graphs connected by linking corresponding vertices like a ladder. This inspires us to make the following definition.

      \begin{description}
	    \item[Definition 51] Consider a graph $\Gamma$ that has $n$ pairs of vertices denoted by $u_{i}$ and $v_{i}$, $i\in\{1,2,\dots,n\}$. Denote the induced subgraph $u_{1}u_{2}\dots u_{n}$ as $\mathcal{U}$, while the induced subgraph $v_{1}v_{2}\dots v_{n}$ as $\mathcal{V}$. The graph $\Gamma$ is a \emph{complete ladder graph} if it satisfies all the following conditions:
        \begin{enumerate}
          \item $\mathcal{U}$ is isomorphic to $K_{n}$;
          \item $\mathcal{V}$ is isomorphic to $K_{n}$;
          \item $\forall i\in[1,n], u_{i}v_{i}\in E(\Gamma)$;
          \item no other edges included in $\Gamma$.
        \end{enumerate}

        \item[Notation 52] $\forall n\in\mathbb{Z}_{+}$, Set
          \begin{center}
            $KL_{n}=\{\Gamma||\Gamma|=2n$ and $\Gamma$ is isomorphic
            to a complete ladder graph$\}$
          \end{center}
      \end{description}

\begin{figure}
      	\centering
        \includegraphics[height=4cm]{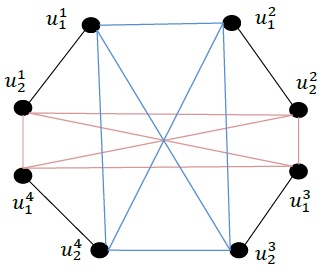}
        \caption{An example of complete ladder graph in $KL_{4}$. Two $K_{4}$s are represented in blue edges and pink edges respectively. }
        \label{Figure 17}
      \end{figure}

      We show an intuitive example of a complete ladder graph in Figure \ref{Figure 17}. Finally, from the proof of Theorem 50, we have
      \begin{description}
	    \item[Corollary 53] $\forall n\in\mathbb{Z}_{+}, KL_{n}\subseteq ST_{n}(n)$.
      \end{description}

    \subsection{Two Further Conclusions about Chromatic Stability}
      In the previous section, we know that the CSI of a stereotype graph $\Gamma_{n}$ ranges from 2 to $n$. Moreover, 2 and $n$ are always the infimum and supremum of $\chi(\Gamma_{n})$ for $n\ge2$. Yet the range is still not completely precise. We need to check whether $\chi(\Gamma_{n})$ can always be any of the positive integers in $[2,n]$ as before.
      \begin{description}
	    \item[Theorem 54] $\forall n\ge2$ and $k\in[2,n], ST_{n}(k)\ne\emptyset$.
      \end{description}

      \begin{proof}
        Let $u_{1}^{i}$ and $u_{2}^{i}$ $(i\in\{1,2,\dots,m\})$ denote $m$ pairs of vertices of any stereotype graph $\Gamma_{m}\in ST_{m}$ $(m\ge2)$. Denote its induced subgraph $u_{1}^{1}u_{2}^{1}u_{1}^{2}u_{2}^{2}\dots u_{1}^{m-1}u_{2}^{m-1}$ as $\Gamma_{m-1}$. Since $\Gamma_{m}\in ST_{m}$, we have $\Gamma_{m-1}\in ST_{m-1}$.

        Let $\Delta=\chi(\Gamma_{m})-\chi(\Gamma_{m-1})$. Since $|V(\Gamma_{m})|-|V(\Gamma_{m-1})|=2$, $\Delta$ can only be 0, 1 or 2. Now suppose $\Gamma_{m-1}$ is given and its edges will not change forever. We discover whether $\Delta$ can be 0,1 or 2 when $\Gamma_{m-1}$ expands to $\Gamma_{m}$ by adding a new pair of vertices $u_{1}^{m}$ and $u_{2}^{m}$.

        A vertex coloring is denoted by a mapping $\theta:V(\Gamma_{m})\rightarrow\{1,2,\dots,w\}$ for some $w$ where $\theta(u_{p}^{i})\ne\theta(u_{q}^{j})$ for any $u_{p}^{i}u_{q}^{j}\in E(\Gamma_{m})$.

        \emph{Case I:} $\Delta=0$.

        This indicates that $\Gamma_{m}$ should be $\chi(\Gamma_{m-1})$-colorable. Let $\theta(u_{1}^{m})=1$ and $\theta(u_{2}^{m})=2$. Then for every $i\in[1,m-1]$,
        \begin{enumerate}
          \item If $\exists p$ s.t. $\theta(u_{p}^{i})=1$, let $u_{p}^{i}u_{2}^{m}\in E(\Gamma_{m})$ and $u_{3-p}^{i}u_{1}^{m}\in E(\Gamma_{m})$;
          \item If $\exists q$ s.t. $\theta(u_{q}^{i})=2$, let $u_{3-q}^{i}u_{2}^{m}\in E(\Gamma_{m})$ and $u_{q}^{i}u_{1}^{m}\in E(\Gamma_{m})$;
          \item Otherwise, whatever type of match is acceptable.
        \end{enumerate}

        We can easily show that the graph $\Gamma_{m}$ we have is not only a stereotype graph but also $\chi(\Gamma_{m-1})$-colorable, i.e. $\Delta=0$.

        \emph{Case II:} $\Delta=1$.

        This indicates that $\Gamma_{m}$ is not $\chi(\Gamma_{m-1} )$-colorable but $(\chi(\Gamma_{m-1})+1)$-colorable. That is, just one of $u_{1}^{m}$ and $u_{2}^{m}$ has to be colored by a new color. We denote such vertex and new color as $\theta(u_{p}^{m})=d$ where $p\in\{1,2\}$ and $d$ is a new color that never occurs in $\Gamma_{m-1}$. Since $u_{p}^{m}u_{3-p}^{m}\in E(\Gamma_{m}), \theta(u_{3-p}^{m})\ne d$.

        Here we expect to have a $K_{\chi(\Gamma_{m-1})+1}$ where $u_{p}^{m}$ is one of its vertices. Suppose its other vertices are $u_{p_{j}}^{m_{j}}$ for $j\in\{1,2,\dots,\chi(\Gamma_{m-1})\}$.

        By Theorem 48, $\chi(\Gamma_{m-1})\le m-1$. Since $u_{p_{j}}^{m_{j}}u_{p}^{m}\in E(\Gamma_{m})$, we have $u_{3-p_{j}}^{m_{j}}u_{3-p}^{m}\in E(\Gamma_{m})$ for $j\in\{1,2,\dots,\chi(\Gamma_{m-1})\}$. Let $\Pi$ denotes the induced subgraph $u_{3-p_{1}}^{m_{1}}u_{3-p_{2}}^{m_{2}}\dots u_{3-p_{t}}^{m_{t}}$, where $t=\chi(\Gamma_{m-1})$.

        If $\Pi$ is colored by less than $t$ colors, let $\theta(u_{3-p}^{m})$ be the different number from $d$ and $u_{3-p_{j}}^{m_{j}}$ for all $j\in\{1,2,\dots,t\}$. Then for each remaining pair of vertices (if exists), whose pair index is $k$, let $u_{3-q}^{k}u_{3-p}^{m}\in E(\Gamma_{m})$ if $\theta(u_{q}^{k})=\theta(u_{3-p}^{m})$ or $q=1$, and then let $u_{q}^{k}u_{p}^{m}\in E(\Gamma_{m})$. The stereotype graph $\Gamma_{m}$ we have now actually satisfies $\chi(\Gamma_{m})=t+1=\chi(\Gamma_{m-1})+1$.

        If $\Pi$ is just colored by $t$ colors, let $\theta(u_{3-p}^{m})=\theta(u_{3-p_{1}}^{m_{1}})$ and then change $\theta(u_{3-p_{1}}^{m_{1}})$ to $d$. The change of $\theta(u_{3-p_{1}}^{m_{1}})$ doesn't change the rationality of the vertex coloring due to $u_{3-p_{1}}^{m_{1}}u_{p}^{m}\notin E(\Gamma_{m})$. Moreover, $\theta(u_{3-p}^{m})$ is now different to $d$. Through the similar procedure above, we again have a stereotype graph $\Gamma_{m}$ satisfying $\chi(\Gamma_{m})=\chi(\Gamma_{m-1})+1$.

        In addition, after we add a pair of vertices in Case II, the new stereotype graph $\Gamma_{m}$ will always include a $K_{\chi(\Gamma_{m})}$. This ensures the rationality of adding a new pair of vertices on $\Gamma_{m}$ and some edges in Case II.

        \emph{Case III:} $\Delta=2$.

        Using a similar approach used in Case II, we can show that Case III can never occur.

        \quad

        With all above, $\Delta$ can only be 0 or 1 for any $\Gamma_{m-1}$ and correlated $\Gamma_{m}$.

        Finally, we illustrate how $K_{2,2}$ can be expanded to a graph in $ST_{n}(k)$ for any $n\ge2$ and $k\in[2,n]$ by adding pairs of vertices and some edges repeatedly.
        \begin{enumerate}
          \item  Through $n-k$ times of adding a pair of vertices in Case I, $K_{2,2}$ is expanded to a graph, denoted $\Gamma^{(I)}$, in $ST_{n-k+2}(2)$.
          \item  Then, through $k-2$ times of adding a pair of vertices in Case II, $\Gamma^{(I)}$ is expanded to a graph in $ST_{n}(k)$.
        \end{enumerate}

        \qedhere
      \end{proof}

      We have already shown that $K_{n,n}\in ST_{n}(2)$ and $KL_{n}\in ST_{n}(n)$, whereas we still don't know the ``structure'' of $ST_{n}(2)$ and $ST_{n}(n)$ so far. The following two theorems show that $K_{n,n}$ is the only graph in $ST_{n}(2)$, and $KL_{n}$ is the only graph in $ST_{n}(n)$.

      \begin{description}
        \item[Theorem 55] $\Gamma_{n}\in ST_{n}(2)\Leftrightarrow\Gamma_{n}$ is isomorphic to $K_{n,n}$.
      \end{description}

      \begin{proof}
        This can be derived from Theorem 47 and the bipartite criterion.

        \qedhere
      \end{proof}

      \begin{description}
        \item[Theorem 56] $\forall n\ge2$, we have
          \begin{equation*}
            \Gamma_{n}\in ST_{n}(n)\Leftrightarrow\Gamma_{n}\mbox{ is isomorphic to }KL_{n}.
          \end{equation*}
      \end{description}

      \begin{proof}
        For $n=2$, by Theorem 55 we have ``$\Gamma_{2}\in ST_{2}(2)$'' $\Leftrightarrow$ ``$\Gamma_{2}$ is isomorphic to $K_{2,2}$'' $\Leftrightarrow$ ``$\Gamma_{2}$ is isomorphic to $KL_{2}$''.

        Now suppose the statement ``$\Gamma_{m}\in ST_{m}(m)\Leftrightarrow\Gamma_{m}$ is isomorphic to $KL_{m}$'' is true for a specific $m\ge2$.

        For any $\Gamma_{m+1}\in ST_{m+1}(m+1)$, let $u_{1}^{i}$ and $u_{2}^{i}$ denote its $m+1$ pairs of vertices $(i\in{1,2,\dots,m+1})$. Let $\Gamma_{m}$ denote its induced subgraph $u_{1}^1u_{2}^{1}u_{1}^{2}u_{2}^{2}\dots u_{1}^{m}u_{2}^{m}$, which belongs to $ST_{m}$.

        From the proof of Theorem 54, $\Delta=\chi(\Gamma_{m+1})-\chi(\Gamma_{m})=0$ or 1. If $\Delta=0$, we have $\chi(\Gamma_{m})=\chi(\Gamma_{m+1})=m+1>m$, which contradicts Theorem 48.

        Thus $\Delta=1$, i.e. $\chi(\Gamma_{m})=m$. Then by the hypothesis, $\Gamma_{m}$ is isomorphic to $KL_{m}$. Let $\mathcal{U}_{0}$ denote its induced graph $u_{p_{1}}^{1}u_{p_{2}}^{2}\dots u_{p_{m}}^{m}$, which is a $K_{m}$. Let $\mathcal{V}_{0}$ denote another induced graph $u_{3-p_{1}}^{1}u_{3-p_{2}}^{2}\dots u_{3-p_{m}}^{m}$, which is also a $K_{m}$.

        Since $\chi(\Gamma_{m+1})=m+1$, one of the $u_{1}^{m+1}$ and $u_{2}^{m+1}$, denoted as $u_{p}^{m+1}$, has to be colored by a new color which doesn't occur in $\Gamma_{m}$. This indicates that there must exist a $K_{m+1}$, denoted as $\mathcal{U}$, where $u_{p}^{m+1}$ is one of its vertices. We label its other vertices as $u_{p_{j}}^{j}$ for $j\in\{1,2,\dots,m\}$. Since $u_{p_{j}}^{j}u_{p}^{m+1}\in E(\Gamma_{m+1})$, we have $u_{3-p_{j}}^{j}u_{3-p}^{m+1}\in E(\Gamma_{m+1})$ for $j\in\{1,2,\dots,m\}$. Denote the induced subgraph $u_{3-p_{1}}^{1}u_{3-p_{2}}^{2}\dots u_{3-p_{m}}^{m}u_{3-{p}}^{m+1}$ as $\mathcal{V}$. We can easily show that $\mathcal{V}$ is also a $K_{m+1}$.

        From the discussion above, we have $\Gamma_{m+1}$ is isomorphic to $KL_{m+1}$ for any $\Gamma_{m+1}\in ST_{m+1}(m+1)$. In addition, when we replace $n$ by $m+1$ in Corollary 53 $\Gamma_{m+1}$ is isomorphic to $KL_{m+1}$.

        Thus ``$\Gamma_{m+1}\in ST_{m+1}(m+1)\Leftrightarrow\Gamma_{m+1}$ is isomorphic to $KL_{m+1}$''. The statement is true for every $n\ge2$ due to mathematical induction.

        \qedhere
      \end{proof}

  \section{Discussion}
    In this article, we raise \emph{stereotype graph} as a theoretical framework in order to describe human's \emph{category stereotypes}. Specifically, we use \emph{Vertices} to represent \emph{labels}, as well as \emph{edges} to represent whether corresponding labels are contradictory or complementary. We also provides CSI, which refers to the chromatic number of a certain graph, as a scale to quantify how stable a certain stereotype lasts in one's cognition. In this way, the framework provide a linkage between mathematical psychology and social cognition.
    
    Our framework is limited to the level of category stereotype. Any other kind of stereotypical thoughts, like ``The thought of committing suicide is pessimistic'' or ``The behavior of committing suicide is stupid.'', are beyond the framework since the so-called ``vertex coloring'' characterizes a person as a member of the group with specific labels. Thus, what stereotype graphs provide is the information of a scheme that someone use to judge others' characteristics rather than others' behaviors or thoughts directly.
    
    \paragraph{More Usual Types of Stereotype Graphs}
      All stereotype graphs we discussed so far are based on Definition 3(3), which forces those graphs to precisely hold six edges among every two pairs of labels which have the same pattern as Figure \ref{Figure 4} or Figure \ref{Figure 5}. Nevertheless, this may not always be the case. For instance, Anne may regard ``Femininity'' and ``Severity'' as contradictory labels while ``Femininity'' and ``Vanity'' are not contradictory at all.

      A more common stereotype graph may have less edges than those stereotype graphs defined by Definition 3. If we remove some edges from the latter one, the chromatic number doesn't increase. Therefore, the stereotype graph we discussed before is the most unstable one.

      Another common circumstance is that after removing enough edges, the stereotype graph becomes disconnected. This indicates that there are several pairs of labels that are not related to each other. In this scenario, the new stereotype graph is more stable.

    \paragraph{Stability of Stereotype Graphs and Stereotypes}
      Next, we discuss the implication of the ``stability'' of stereotype graphs. The most stable stereotype graph must have the same structure as shown in Definition 3. But how stable will it be in fact? From Theorem 48, we know that the CSI of a stereotype graph $\Gamma_{n}$ is in $[2,n]$ where $n$ is the number of pairs of contradictory labels. However, pairs of contradictory labels in human cognition may be far less than the number of people anyone knows in their whole life.

      This may explain why most people maintain, more or less, stereotypes. Under most circumstances, stereotypes can be stable in human cognition at least for a while. The fewer pairs of contradictory labels, the more stable a stereotype will be in human cognition. Therefore, no matter whether a stereotype graph is based on Definition 3 or not would still seem ``stable'' in human cognition. This is a manifestation of the stereotype effect and the resistance of the stereotype to change itself.

    \paragraph{Innovation and Prospect of Stereotype Graphs}
      We use graph theory to discuss stereotypes, which is a common subject in the field of social psychology. The usage of graph theory numericalizes the stereotype as a mathematical concept. We believe that it can provide a new perspective and thought on studying stereotypes and social interaction in psychology.

      Our research is not limited to psychology. In artificial intelligence (AI), for instance, we can enable a robot to cognize and categorize human personality in the sense of human intelligence with the help of stereotype graphs. This is a kind of simulation of human cognition, as stereotypes help humans to make sense of the social world, potentially enlightening cognition-inspired AI for multi-agent interaction.

  \section{Conclusions}
    We use \emph{stereotype graphs} as a mathematical framework to portray \emph{category stereotypes} in human's cognition. By using methods and results in graph theory, we provide six criteria for judging the \emph{stability} of a given stereotype based on complementary labels. In comparison to the former three criteria, the latter three criteria are more convenient to use computationally. We establish \emph{chromatic stability index} (CSI) to evaluate the stability of a contradictory-labels-based stereotype. We provide a range of CSI with proof. We also show the uniqueness of the structures of the most stable and unstable stereotype graphs. Since the pairs of contradictory labels are usually not too many in human cognition, most people would easily have, more or less, stereotypes. Stereotypes also show certain resistance to change. The innovative mathematical framework of stereotype graphs in this article can be applied to other fields, such as cognition-inspired AI for multi-agent interaction.

  \section*{Acknowledgements}
    This paper was finished and first presented in 2016 for the 8th Science Innovation of the School of Mathematical Sciences at Fudan University. The research was supported by the corresponding grant from the School of Mathematical Sciences at Fudan University. The funders had no role in study design, decision to publish, or preparation of the manuscript. The authors have declared that no competing interests exist. The author sincerely thanks Prof. Xiuli He and Dr. Tianshu Feng for their great help in revising the draft and providing precious suggestions. The author extends special thanks to Prof. Shenglin Zhu for his tremendous help on reading early drafts and raising valuable suggestions during early supervision. The author would also like to thank Xiao Ren, Huachun Zhu, and Tom-Clarence Yan for providing constructive opinions on the study before the paper writing.
  
  \bibliographystyle{abbrvnat}
  \bibliography{manuscript}

\end{document}